\documentclass[11pt]{article}

\usepackage[a4paper,margin=2.5cm]{geometry}
\usepackage[T1]{fontenc}
\usepackage[utf8]{inputenc}
\usepackage{graphicx}
\usepackage{booktabs}
\usepackage{subcaption}
\usepackage{amsmath}
\usepackage{tikz}
\usepackage{url}
\usepackage{authblk}
\usepackage[numbers,super,sort&compress]{natbib}
\usepackage[hidelinks]{hyperref}

\newcounter{suppnote}
\newcommand{\suppnote}[1]{%
  \refstepcounter{suppnote}%
  \section*{Supplementary Note \thesuppnote\ --- #1}%
}

\title{\bfseries Integrating GNSS-Derived Zenith Wet Delay into a Weather Foundation Model Improves Precipitation Forecasting}

\author[1]{Leonardo Trentini\thanks{Corresponding author: \texttt{ltrentini@ethz.ch}}}
\author[2,3]{Fanny Lehmann}
\author[1]{Laura Crocetti}
\author[1,2]{Benedikt Soja}

\affil[1]{Institute of Geodesy and Photogrammetry, ETH Zurich, Switzerland}
\affil[2]{ETH AI Center, ETH Zurich, Switzerland}
\affil[3]{Computational and Applied Mathematics Laboratory, Seminar for Applied Mathematics, ETH Zurich, Switzerland}

\date{}

\begin{document}
\maketitle

\begin{abstract}
\noindent
Global Navigation Satellite Systems (GNSS), best known for positioning, also serve weather science, as atmospheric water vapour delays their signals. This delay, the Zenith Wet Delay (ZWD), is a direct, all-weather measure of column moisture. Although assimilated into numerical weather prediction for decades, ZWD is not yet used by leading Machine Learning Weather Models (MLWM), despite addressing a known deficiency: the underestimation of severe precipitation.
Here we present the first integration of GNSS-derived ZWD into Aurora, a state-of-the-art weather foundation model. Our extended Aurora learns ZWD with skill comparable to its pretrained variables. More importantly, including ZWD systematically improves forecasts when fine-tuning for 6-hour accumulated precipitation. Gains grow with severity, reaching an 8.8\% increase in Equitable Threat Score at the 99th percentile, while the precipitation power spectrum becomes more realistic at synoptic and planetary scales. GNSS observations therefore encode information that MLWM can exploit for high-impact precipitation.
\end{abstract}

\section{Introduction}

The past years have witnessed a transformation in atmospheric forecasting driven by deep learning. Data-driven models such as FourCastNet\citep{pathak2022}, Pangu-Weather\citep{bi2023}, GraphCast\citep{lam2023} and the probabilistic GenCast\citep{price2025}, among others, have demonstrated global medium-range forecast skill competitive with operational Numerical Weather Prediction (NWP) at a fraction of the computational cost\citep{rasp2024}. More recently, the paradigm has shifted toward weather \textit{foundation models}: large systems pretrained on heterogeneous atmospheric datasets that can be fine-tuned for specific downstream tasks\citep{nguyen2023, Bodnar2025, ozdemir2026esfm, schmude2024prithviwxcfoundationmodel}. Aurora\citep{Bodnar2025}, among the most capable of these, achieves state-of-the-art performance on standard benchmark variables while its Perceiver-based encoder architecture is designed to accommodate new data types without requiring modifications to the core architecture. Yet, despite this architectural flexibility, most existing weather foundation models are pretrained primarily on gridded reanalysis products such as ERA5\citep{hersbach2020} and do not incorporate in-situ observations from geodetic networks. A parallel line of work is broadening the data foundation of these models, either by extending model encoders to heterogeneous observational data\citep{ozdemir2026esfm} or by learning end-to-end from raw observations rather than reanalysis\citep{vaughan2025}, underscoring the value of bringing observational streams directly into machine learning weather pipelines.

Ground-based Global Navigation Satellite System (GNSS) receivers, beyond their original purpose of providing accurate localization, can provide a unique source of atmospheric information currently unused by leading machine learning weather models. As a GNSS signal propagates through the troposphere it accumulates a path delay proportional to the integral of refractivity along the line of sight. Mapping this delay to the zenith direction and removing the hydrostatic contribution isolates the Zenith Wet Delay (ZWD), a vertical integral of moist refractivity that depends almost entirely on water vapour. ZWD maps linearly to the Integrated Water Vapour through $\mathrm{IWV} = \Pi(T_m)\,\mathrm{ZWD}$, where the dimensionless coefficient $\Pi$ varies only weakly with the column-mean atmospheric temperature $T_m$\citep{bevis1992, bevis1994}, making ZWD a direct, weather-independent observable of column water vapour, validated against radiosondes and reanalyses\citep{wang2008}. Growing global networks now comprise thousands of continuously operating GNSS stations\citep{yuan2023, blewitt2018}, providing sub-hourly ZWD estimates under all sky conditions, including during active precipitation. GNSS ZWD has been assimilated into operational NWP for several decades\citep{vedel2004, guerova2016, jones2019}, with documented benefits for high-impact rainfall events\citep{giannaros2020}, and used to monitor phenomena ranging from Alpine Foehn\citep{aichinger2022} and the El Ni\~no Southern Oscillation\citep{crocetti2024b} to convective storms\citep{aichinger2023b} and typhoons\citep{zhu2020}. While GNSS-derived tropospheric delays have themselves been the target of machine learning and foundation models \citep{li2024, zhang2022transformer, ding2024}, the inverse direction, feeding GNSS-derived ZWD into a state-of-the-art atmospheric foundation model, has not been attempted to our knowledge. ZWD is moreover not represented as a state variable in ERA5.

This omission is particularly consequential for precipitation forecasting: heavy precipitation drives some of the most damaging weather hazards while remaining among the hardest forecast targets for both numerical and machine learning weather models\citep{olivetti2024}. Precipitation is intermittent, strongly non-Gaussian, represented with substantial bias in ERA5\citep{lavers2022}, and strongly modulated by the fine-scale spatial and temporal distribution of column moisture\citep{lam2023, pasche2025, zhang2025}. Because reanalysis humidity fields are smoothed by the assimilation cycle, foundation models trained on them inherit a filtered water-vapour representation and underestimate precipitation variability at convective scales, and independent verification against GNSS networks has indeed revealed systematic water vapour biases across multiple AI weather models\citep{ding2025}. GNSS ZWD, as a direct, unfiltered integral of wet refractivity, captures column-moisture variability tied to the moisture-flux convergence and convective triggering preceding heavy rainfall: a complementary, physically grounded observable that reanalysis humidity fields partly smooth out.

\begin{figure}[!ht]
\centering
\includegraphics[width=\linewidth]{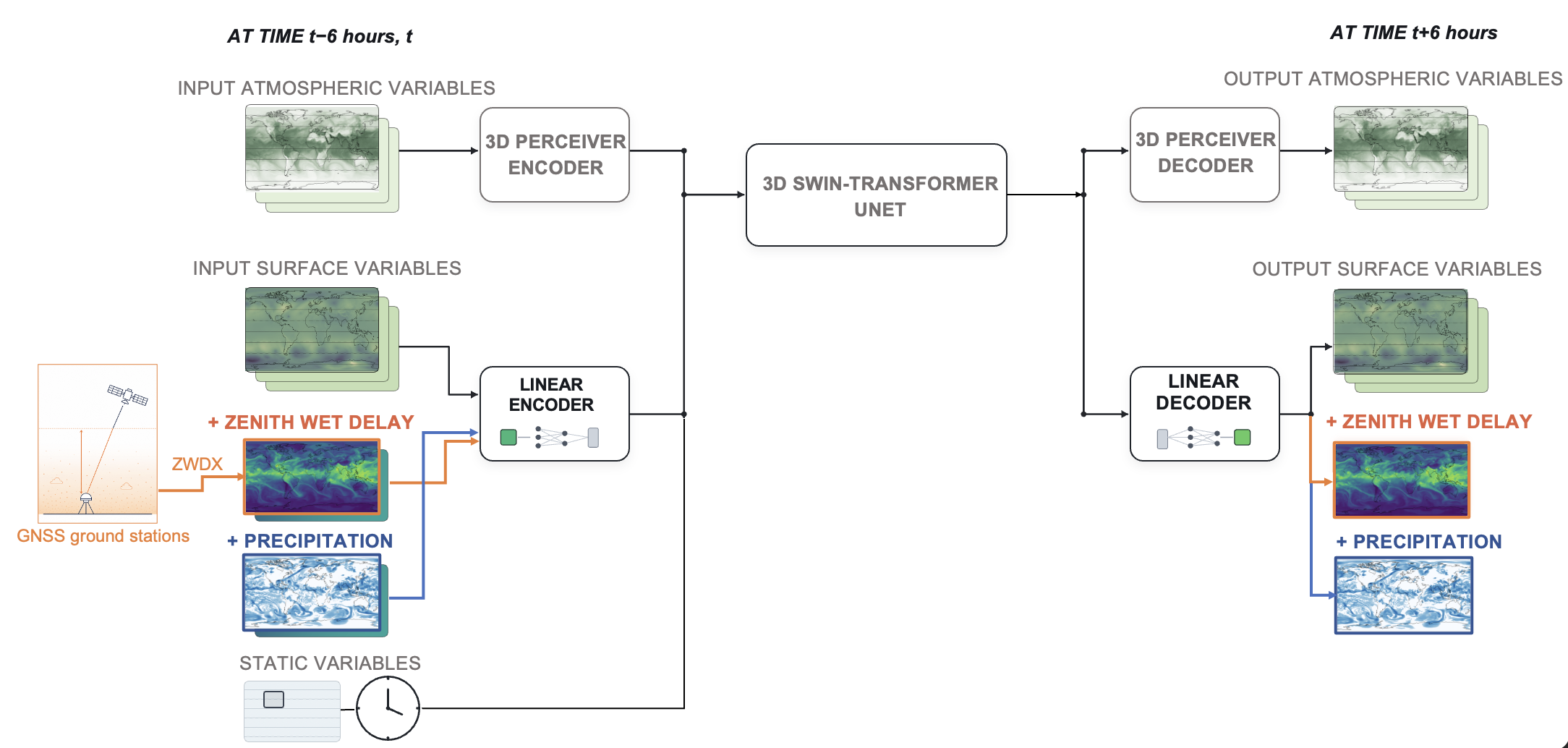}
\caption{\textbf{Integrating a GNSS-derived observable into a weather foundation model.} Ground-based GNSS receivers measure the tropospheric delay accumulated by satellite signals; removing the hydrostatic contribution isolates the Zenith Wet Delay (ZWD), a direct, all-weather measure of column water vapour, which the ZWDX product renders as a global $0.25^{\circ}$ field. ZWD and 6-hour accumulated precipitation - both absent from Aurora's pretraining - are added as new surface variables (highlighted) and enter through an extension of the linear surface encoder and decoder alone; the Perceiver-based encoder, 3D Swin Transformer U-Net backbone and decoder are inherited from the pretrained checkpoint. The baseline architecture is identical but for the removal of ZWD, so the contrast between the two isolates its contribution.}
\label{fig:overview}
\end{figure}

Here we ask whether supplying this observable to a weather foundation model measurably reduces that deficiency. Using ZWDX\citep{crocetti2024}, a global gridded ZWD product produced by an XGBoost model trained on over 19{,}000 GNSS stations, we extend Aurora with ZWD as a new surface variable and fine-tune it for 6-hour accumulated precipitation - a variable absent from pretraining - with and without ZWD as an auxiliary input (Figure~\ref{fig:overview}). Because the two precipitation models share initialisation, optimiser, schedule, step budget and training data and differ only in their variable set, the contrast between them isolates the contribution of GNSS ZWD to forecast skill. We find that ZWD is learned at a skill level on par with Aurora's already-pretrained variables; that including it improves precipitation skill systematically, with gains that grow monotonically with event severity; and that the predicted precipitation power spectrum becomes markedly more realistic across all resolved scales. To our knowledge this is the first integration of a geodetic observable into an atmospheric foundation model, and it establishes that observing systems outside the reanalysis stack carry information that this emerging generation of models can exploit on precisely their hardest forecast target.

\section{Results}

\subsection{Fine-tuning brings ZWD to parity with already-pretrained variables}
\label{sec:zwd_learning}

To assess whether Aurora can accurately predict ZWD as an auxiliary output, we compare its Step~1 skill on ZWD against the four standard surface variables seen throughout pretraining (10\,m zonal and meridional wind components \textit{u10} and \textit{v10}, 2\,m air temperature \textit{t2m}, and mean sea-level pressure \textit{MSLP}) and against specific humidity ($q$), the pressure-level moisture quantity that is ZWD's most direct counterpart. All variables in Table~\ref{tab:zwd_skill} are evaluated on the same Step~1 fine-tuned model on the held-out test set at 6-hour lead time, isolating how well a previously unseen variable is learned relative to a brief refinement of variables already pretrained.

ZWD is learned essentially at parity with specific humidity ($R = 0.998$ for both, FSS$_{95}$ = 0.985 vs 0.986), the level-resolved quantity from which the column integral can in principle be derived. In physical units it is predicted with an MAE of $3.03$\,mm and an RMSE of $5.23$\,mm, and its relative MAE of $1.92\%$ (Table~\ref{tab:zwd_skill}) is of the same order as those of the pretrained surface fields. Its ETS (Table~\ref{tab:zwd_skill}, right) is competitive at moderate thresholds (75th-95th percentile) and degrades more steeply only at the 99th percentile, as expected from the rarity of extreme column-moisture-loading events, while its spectral skill lies well within the envelope of the pretrained variables (integrated LSD $0.252$, against a range of $0.113$ for MSLP to $0.339$ for u10; see the band-wise breakdown in Supplementary Note~7). ZWD is thus learned to comparable skill through Aurora's standard variable-embedding mechanism, without any bespoke modules, dedicated prediction heads, or structural modifications to the encoder, backbone, or decoder.

\begin{table}[!ht]
\centering
\caption{\textbf{Step~1 deterministic and threshold-based skill of ZWD compared to Aurora's standard pretrained variables.} Standard pretrained variables taken into account are u10, v10, t2m, MSLP and specific humidity ($q$), averaged over all 13 pressure levels. All variables are evaluated after the same Step~1 fine-tuning on the held-out test set at 6-hour lead time. Arrows indicate whether lower or higher is better.}
\label{tab:zwd_skill}
\small
\setlength{\tabcolsep}{5pt}
\begin{tabular}{l|ccc|cccc}
\toprule
& \multicolumn{3}{c|}{Deterministic} & \multicolumn{4}{c}{\quad Equitable Threat Score (ETS)} \\
\cmidrule(lr){2-4}\cmidrule(lr){5-8}
Var. & $R$ $\uparrow$ & FSS$_{95}$ $\uparrow$ & Rel.\ MAE $\downarrow$ & ETS$_{75}$ $\uparrow$ & ETS$_{90}$ $\uparrow$ & ETS$_{95}$ $\uparrow$ & ETS$_{99}$ $\uparrow$ \\
\midrule
MSLP & 0.9998 & 0.996 & 0.013\% & 0.964 & 0.960 & 0.953 & 0.941 \\
t2m  & 0.9997 & 0.977 & 0.067\% & 0.956 & 0.846 & 0.828 & 0.856 \\
u10  & 0.9968 & 0.992 & 4.72\%  & 0.916 & 0.920 & 0.907 & 0.847 \\
v10  & 0.9956 & 0.989 & 6.06\%  & 0.894 & 0.886 & 0.884 & 0.861 \\
$q$  & 0.9982 & 0.986 & 2.99\%  & 0.938 & 0.935 & 0.889 & 0.708 \\
\midrule
ZWD & 0.9983 & 0.985 & 1.92\% & 0.947 & 0.912 & 0.851 & 0.673 \\
\bottomrule
\end{tabular}
\end{table}

\subsection{ZWD improves precipitation forecasting skill}
\label{sec:precip}

Having established that ZWD is skillfully learned, we now assess whether its inclusion as an auxiliary target improves precipitation forecasting in Aurora. Models A (``With ZWD'') and B (baseline, ``Without ZWD'') are evaluated on the held-out test set using the metric suite defined in Methods. Headline scores are quoted at a 6-hour lead time, as ZWD is a short-lived signature of the instantaneous moisture column and its predictive value is expected to be greatest at short range; how the gain evolves over longer autoregressive rollouts is reported below under \textit{Threshold-based skill}.

Unless otherwise noted, every Step~2 metric reported below is the mean over ten training checkpoints saved at the same training steps for both models on the late-training plateau, quoted with sample standard deviation (s.d.) across those checkpoints where space permits. Because Models A and B share initialisation, optimiser, schedule, step budget and data and differ only in ZWD, checkpoints saved at the same step are matched pairs; we therefore assess the significance of the With-versus-Without-ZWD difference with a checkpoint-wise paired $t$-test across these matched steps, reporting alongside it the number of checkpoints at which ZWD wins (see Methods, ``Checkpoint ensemble and paired significance testing'').

\subsubsection{Deterministic metrics}

Table~\ref{tab:precip_det} reports deterministic skill on the test period. Including ZWD improves the pointwise error metrics: MAE decreases by 0.7\%, RMSE by 1.8\% and MSE by 3.7\%, with the largest gains in MSE-based metrics indicating a particular reduction of large errors. The FSS at the 95th percentile improves by 1.5\% and the Pearson $R$ is essentially unchanged ($-0.2\%$), so the linear correlation of predictions with observations is preserved while their spatial and intensity structure are improved. The precipitation gain is reproduced at 8-9 of the ten checkpoints and is significant under the paired test.

\begin{figure}[!ht]
\centering
\includegraphics[width=0.82\linewidth]{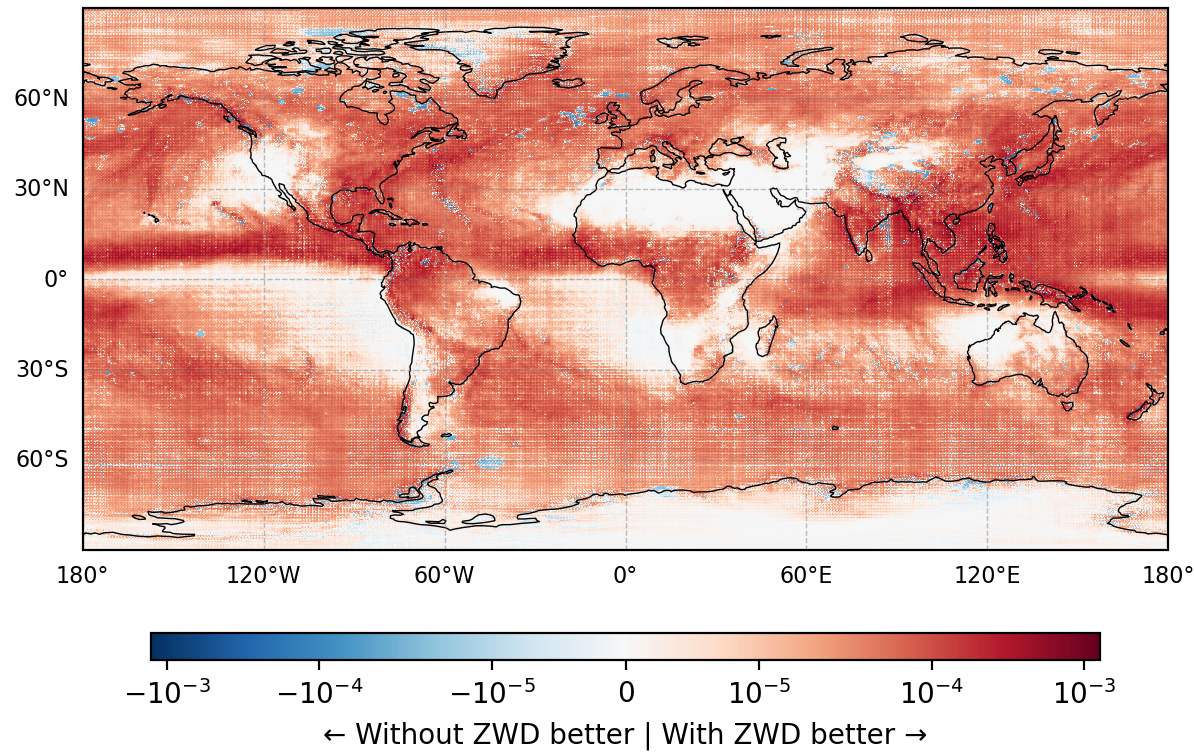}
\caption{\textbf{ZWD reduces 6-hour precipitation RMSE across most of the globe.} Global spatial distribution of the 6-hour precipitation RMSE difference for the main-analysis checkpoint (red: ``With ZWD'' improves upon the baseline ``Without ZWD''; blue: baseline outperforms).}
\label{fig:precip_map}
\end{figure}

The global map of RMSE differences (Figure~\ref{fig:precip_map}) shows that ZWD reduces precipitation RMSE across most of the globe, with the largest gains in the tropics and mid-latitude storm tracks - regions of strong moisture-flux convergence, active deep convection and high ZWD spatial variability, where the column-moisture signal best discriminates raining from non-raining columns. The few regions where the baseline marginally outperforms are spatially confined and generally coincide with arid climatologies in which precipitation is rare and column moisture is uniformly low, so ZWD adds little discriminating information.

\begin{table}[!ht]
\centering
\caption{\textbf{Step~2 deterministic precipitation skill on the held-out test set.} Mean $\pm$ s.d.\ over ten training checkpoints saved at matched training steps, for Model A (``With ZWD'') and Model B (``Without ZWD''); bold marks the better model. The RMSE improvement is significant under the checkpoint-wise paired $t$-test ($9/10$ checkpoints favour ZWD, $p = 0.032$), and every metric improves at 8-9 of the ten checkpoints.}
\label{tab:precip_det}
\small
\setlength{\tabcolsep}{8pt}
\begin{tabular}{lccc}
\toprule
Metric & Without ZWD & With ZWD & Rel.\ $\Delta$ \\
\midrule
MAE [mm]     & $0.195 \pm 0.007$ & $\mathbf{0.193 \pm 0.004}$ & $-0.7\%$ \\
RMSE [mm]    & $1.015 \pm 0.034$ & $\mathbf{0.997 \pm 0.019}$ & $-1.8\%$ \\
MSE [mm$^2$] & $1.032 \pm 0.071$ & $\mathbf{0.994 \pm 0.038}$ & $-3.7\%$ \\
Pearson $R$  & $\mathbf{0.853 \pm 0.001}$ & $0.851 \pm 0.001$ & $-0.2\%$ \\
FSS 95\%     & $0.917 \pm 0.022$ & $\mathbf{0.931 \pm 0.011}$ & $+1.5\%$ \\
\bottomrule
\end{tabular}
\end{table}

\subsubsection{Threshold-based skill}

The ETS scores in Table~\ref{tab:precip_ets} show that the relative benefit of ZWD increases with the precipitation threshold. At the 75th percentile the gain is $+1.2\%$; at the 99th percentile it reaches $+8.8\%$, and all four gains are significant under a checkpoint-wise paired test ($p \leq 0.013$). The advantage is not carried by a handful of outlier events: resolved timestep by timestep across the test period, the ZWD-enriched model leads the baseline at essentially every verification time (see the per-timestep series in Supplementary Note~7). This pattern is physically intuitive: ZWD's largest excursions coincide with the high-moisture conditions that drive heavy precipitation, so it is most informative in the tail of the distribution, where moisture availability is the binding constraint on rainfall.

\begin{table}[!ht]
\centering
\caption{\textbf{Step~2 threshold-based precipitation skill (Equitable Threat Score) on the held-out test set.} Mean $\pm$ s.d.\ over ten training checkpoints saved at matched training steps; bold marks the better model. \textit{ZWD wins} is the number of those checkpoints favouring Model A (``With ZWD''); \textit{paired $p$} is the two-sided checkpoint-wise paired $t$-test across matched steps ($^{*}\,p<0.05$).}
\label{tab:precip_ets}
\small
\setlength{\tabcolsep}{8pt}
\begin{tabular}{lccccc}
\toprule
Threshold & Without ZWD & With ZWD & Rel.\ $\Delta$ & ZWD wins & paired $p$ \\
\midrule
75th percentile & $0.695 \pm 0.016$ & $\mathbf{0.703 \pm 0.009}$ & $+1.2\%$ & 9/10 & $0.013^{*}$ \\
90th percentile & $0.656 \pm 0.022$ & $\mathbf{0.669 \pm 0.012}$ & $+1.9\%$ & 9/10 & $0.012^{*}$ \\
95th percentile & $0.581 \pm 0.036$ & $\mathbf{0.604 \pm 0.020}$ & $+3.9\%$ & 9/10 & $0.007^{*}$ \\
99th percentile & $0.395 \pm 0.052$ & $\mathbf{0.430 \pm 0.031}$ & $+8.8\%$ & 8/10 & $0.004^{*}$ \\
\bottomrule
\end{tabular}
\end{table}

\begin{figure}[!ht]
\centering
\includegraphics[width=0.92\linewidth]{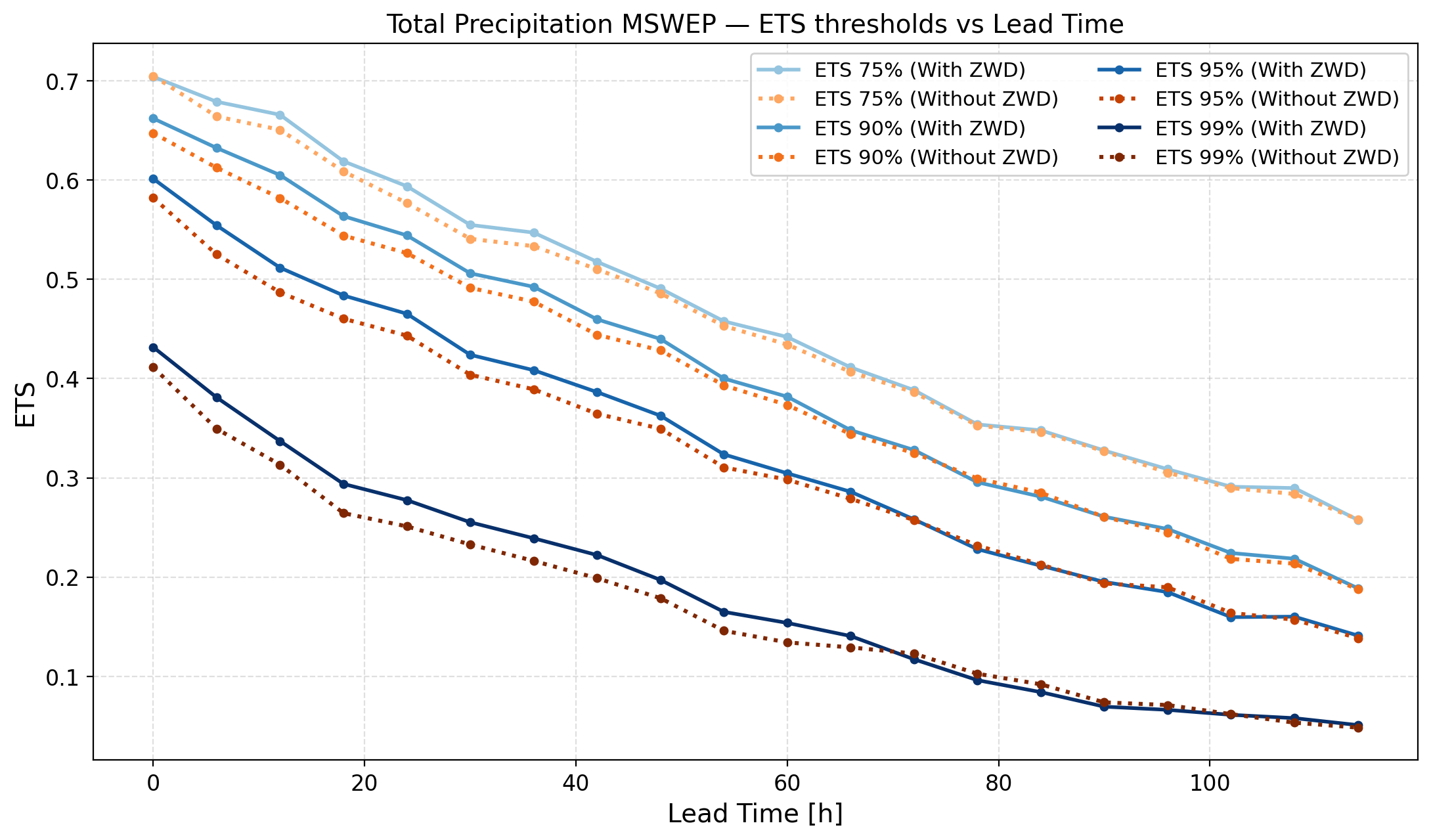}
\caption{\textbf{The ZWD advantage persists through the short-to-intermediate forecast range.} Equitable Threat Score as a function of lead time at four climatological percentiles (75th, 90th, 95th, 99th) for Model~A (``With ZWD''; solid) and Model~B (``Without ZWD''; dotted), averaged over ten initialisation dates spanning the held-out test set. Forecasts are produced purely autoregressively, without rollout fine-tuning. Skill decays smoothly and monotonically for both models at every threshold; the separation is largest at the extreme thresholds and at the shortest leads, and closes beyond ${\approx}70$\,h as both models lose skill.}
\label{fig:precip_ets_lead}
\end{figure}

The gain also persists well beyond the 6-hour lead time at which it is quoted. Evaluated over autoregressive rollouts out to 114\,h - neither model having received any rollout fine-tuning - skill decays smoothly and monotonically at every threshold, with no instability in either model (Figure~\ref{fig:precip_ets_lead}). Model A holds a clear advantage at the 95th and 99th percentiles through the short-to-intermediate range, out to ${\approx}60$-$70$\,h, and at the 75th and 90th it matches the baseline without ever falling below it; beyond ${\approx}70$\,h the curves converge as both models lose skill. The benefit of ZWD is therefore concentrated at early-to-intermediate lead times, rather than confined to the first step. See the corresponding rollout RMSE for precipitation and for the pretrained variables in Supplementary Note~6.

\begin{figure}[!ht]
\centering

\begin{minipage}[t]{0.49\linewidth}
\centering
\begin{tikzpicture}
  \node[anchor=north west,inner sep=0] (img)
    {\includegraphics[width=\linewidth]{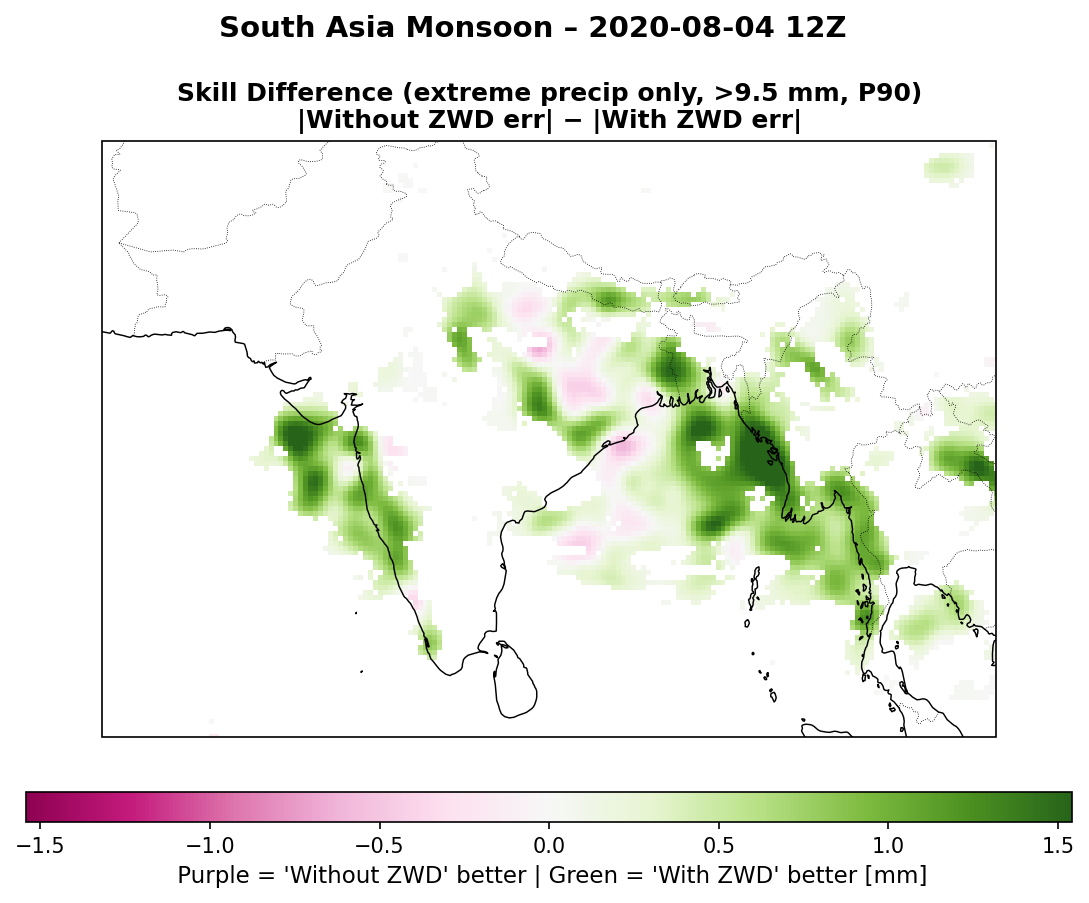}};
  \node[anchor=north west,overlay] at ([xshift=-7pt,yshift=5pt]img.north west) {a};
\end{tikzpicture}
\end{minipage}
\hfill
\begin{minipage}[t]{0.49\linewidth}
\centering
\begin{tikzpicture}
  \node[anchor=north west,inner sep=0] (img)
    {\includegraphics[width=\linewidth]{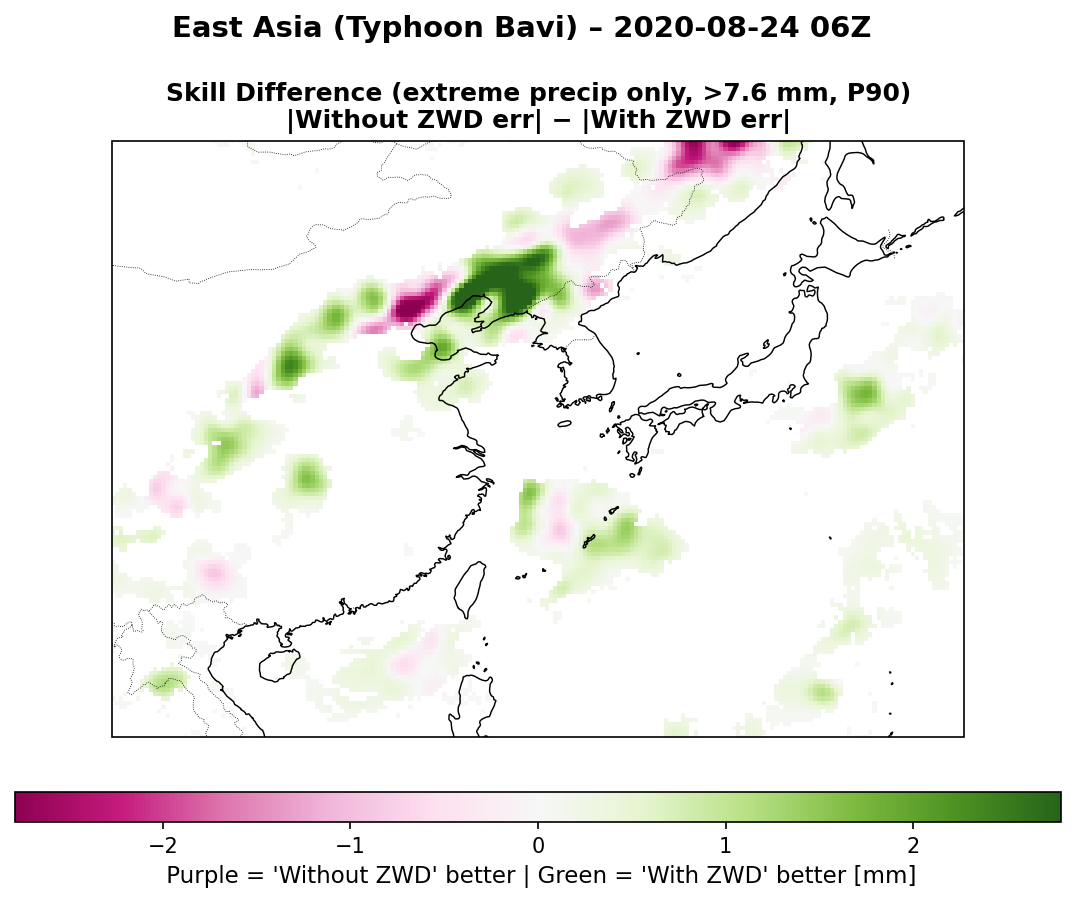}};
  \node[anchor=north west,overlay] at ([xshift=-7pt,yshift=5pt]img.north west) {b};
\end{tikzpicture}
\end{minipage}

\caption{\textbf{Where ZWD improves the forecast during high-impact events.} Per-cell skill difference for extreme precipitation cells only ($>$P90 of the observed regional distribution), computed as the absolute ``Without ZWD'' error minus the absolute ``With ZWD'' error, so that green indicates the ZWD-enriched forecast is closer to MSWEP and purple the baseline: \textbf{a} South Asia monsoon, initialised 2020-08-04 12Z; \textbf{b} Typhoon Bavi, East Asia, 2020-08-24 06Z. Improvement is spatially coherent and concentrated on the heavy-rain structures of each system. The intensity-resolved error reduction is given in Figure~\ref{fig:precip_regional}, and the corresponding forecast fields in Supplementary Note~8.}
\label{fig:precip_event_maps}
\end{figure}

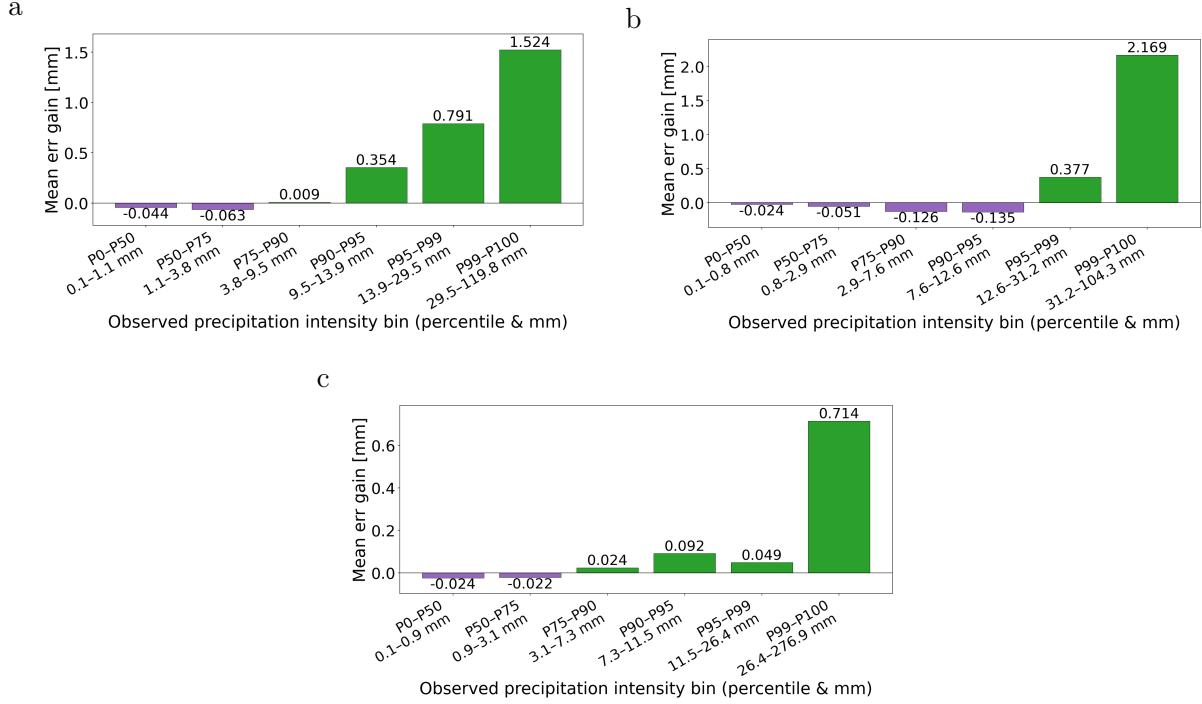
\begin{figure}[!ht]
\centering

\begin{minipage}[t]{0.49\linewidth}
\centering
\begin{tikzpicture}
  \node[anchor=north west,inner sep=0] (img)
    {\includegraphics[width=\linewidth]{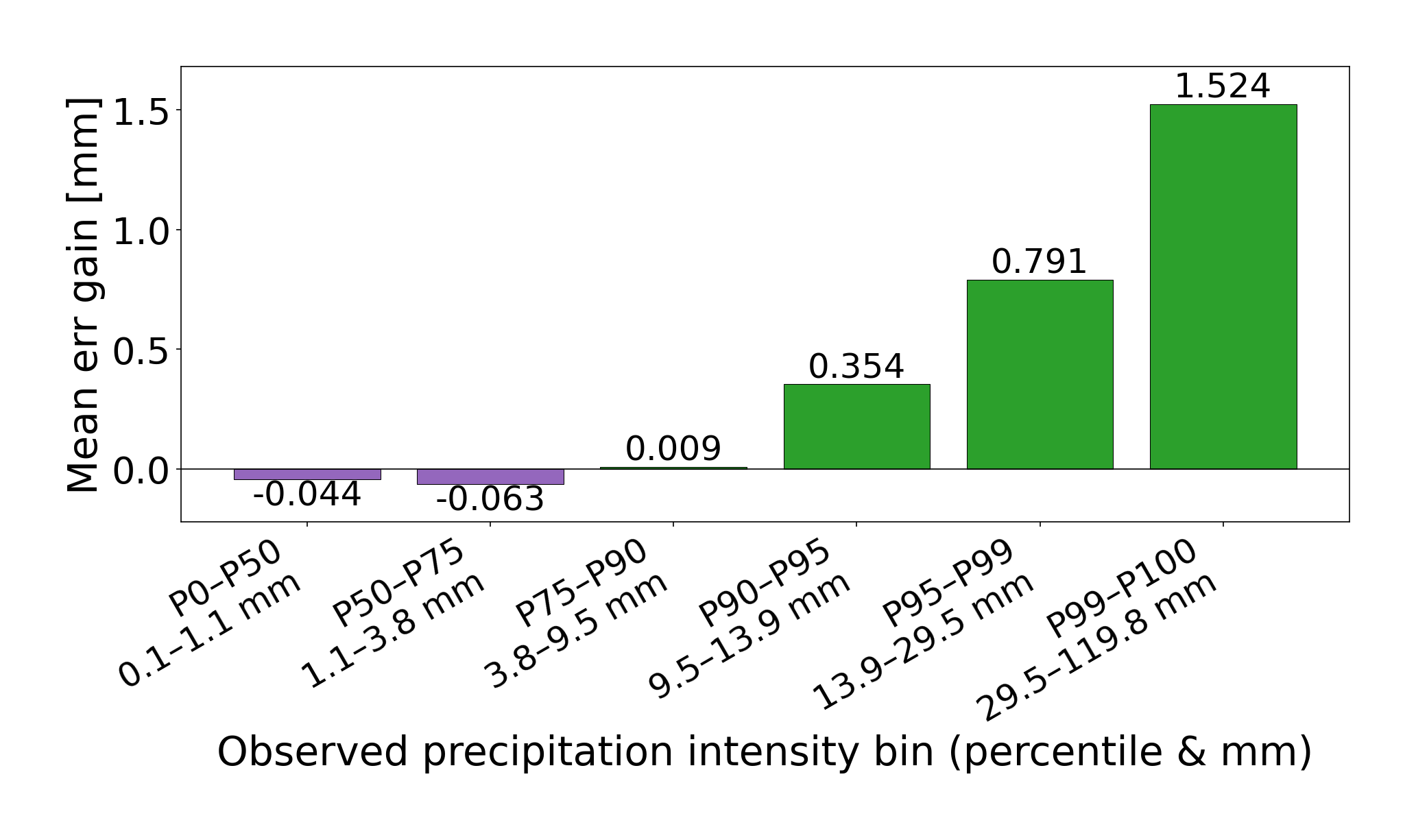}};
  \node[anchor=north west,overlay] at ([xshift=-7pt,yshift=5pt]img.north west) {a};
\end{tikzpicture}
\end{minipage}
\hfill
\begin{minipage}[t]{0.49\linewidth}
\centering
\begin{tikzpicture}
  \node[anchor=north west,inner sep=0] (img)
    {\includegraphics[width=\linewidth]{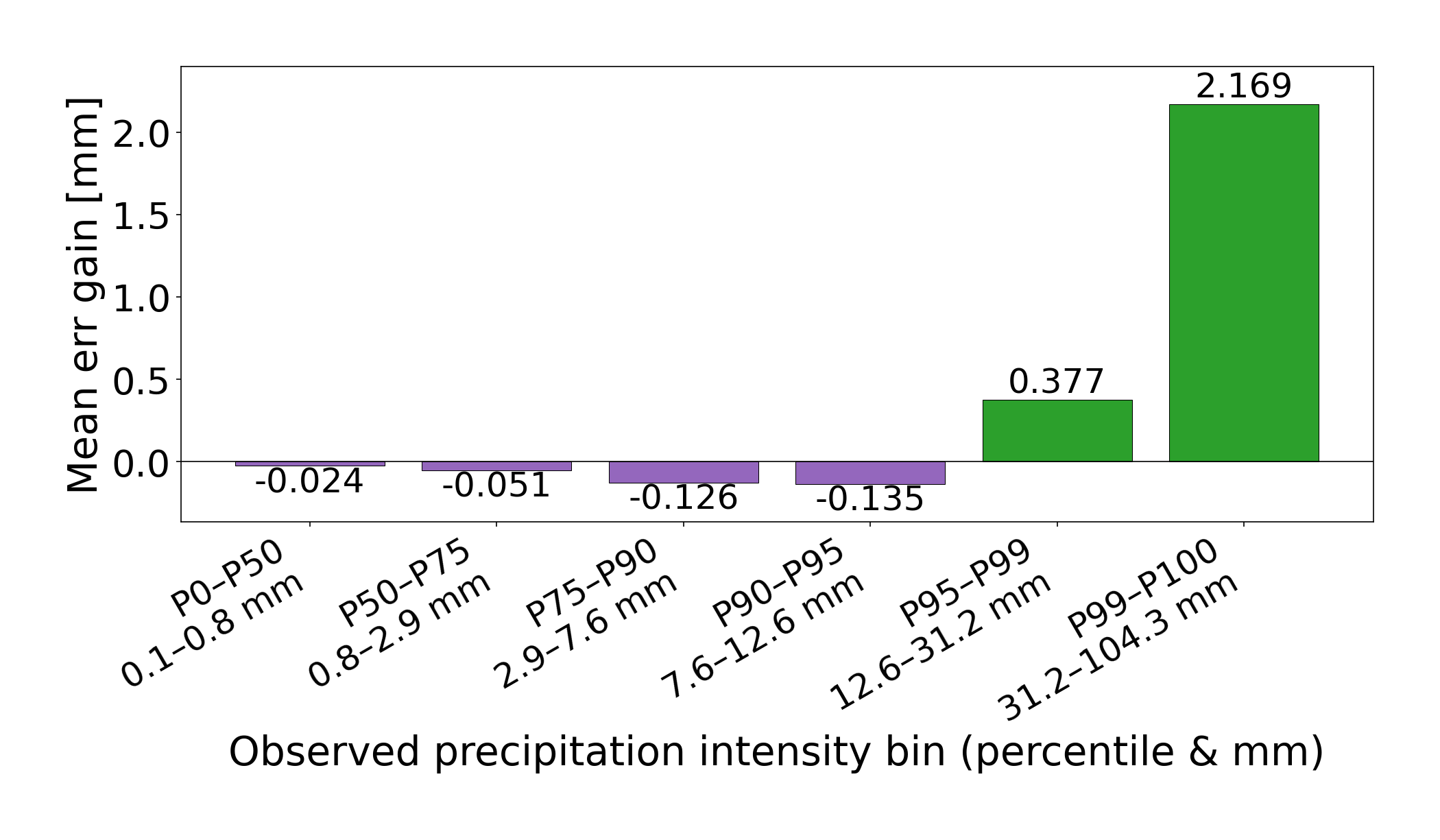}};
  \node[anchor=north west,overlay] at ([xshift=-7pt,yshift=5pt]img.north west) {b};
\end{tikzpicture}
\end{minipage}

\vspace{0.6em}

\begin{minipage}[t]{0.49\linewidth}
\centering
\begin{tikzpicture}
  \node[anchor=north west,inner sep=0] (img)
    {\includegraphics[width=\linewidth]{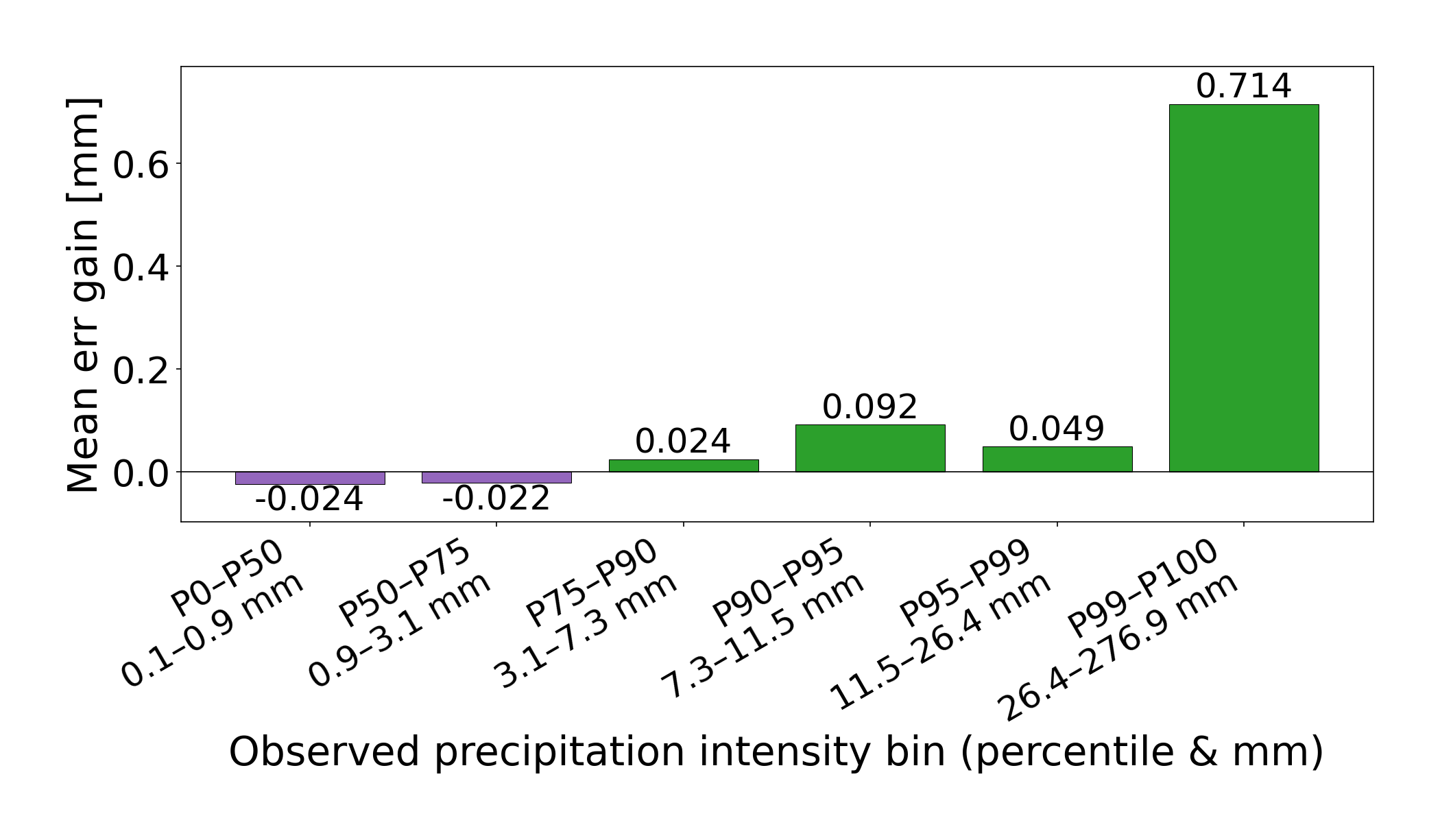}};
  \node[anchor=north west,overlay] at ([xshift=-7pt,yshift=5pt]img.north west) {c};
\end{tikzpicture}
\end{minipage}

\caption{\textbf{The ZWD gain concentrates in the heavy-precipitation tail of individual high-impact events.} Precipitation error reduction (positive $=$ ``With ZWD'' better) versus observed-intensity bin for three extreme events drawn from distinct basins and regimes, all within the held-out test set: \textbf{a} South Asia monsoon (initialised 2020-08-04 12Z), \textbf{b} Typhoon Bavi, East Asia (2020-08-24 06Z), \textbf{c} Hurricane Delta, Central America (2020-10-04 06Z). ZWD is neutral or marginally negative at low-to-moderate intensities and strongly positive in the extreme upper tail in every case. The spatial distribution of the same gain is shown in Figure~\ref{fig:precip_event_maps}; see further regional time series and exceedance diagnostics in Supplementary Note~8.}
\label{fig:precip_regional}
\end{figure}

To confirm that this benefit manifests during individual high-impact events, and not only in the global aggregate, we examine three extreme cases from distinct basins and regimes: an active South Asian monsoon episode, Typhoon Bavi over East Asia, and Hurricane Delta over Central America, all falling within the held-out test set. Mapping the per-cell skill difference over the extreme cells of each event (Figure~\ref{fig:precip_event_maps}) localises where the benefit arises: the monsoon case is overwhelmingly improved over the regions of heaviest rainfall, and the typhoon shows a coherent band of improvement along its principal rain swath. Stratifying the mean precipitation error reduction by observed-precipitation intensity (Figure~\ref{fig:precip_regional}) reveals the same signature in every case: ZWD is neutral or marginally negative at low-to-moderate intensities and strongly positive in the extreme upper tail, reaching a mean error reduction of $+1.52$, $+2.17$ and $+0.71$\,mm in the top intensity bin for the three events, respectively. Even the hurricane, whose bulk regional RMSE is essentially unchanged, retains this tail gain, confirming that the benefit is localized to the heavy-precipitation regime where the column-moisture signal is most informative. See further per-event diagnostics - regional skill time series, per-cell skill differences and threshold-exceedance classifications - in Supplementary Note~8.

\subsubsection{Spectral skill}

\begin{figure}[!ht]
\centering
\includegraphics[width=0.68\linewidth]{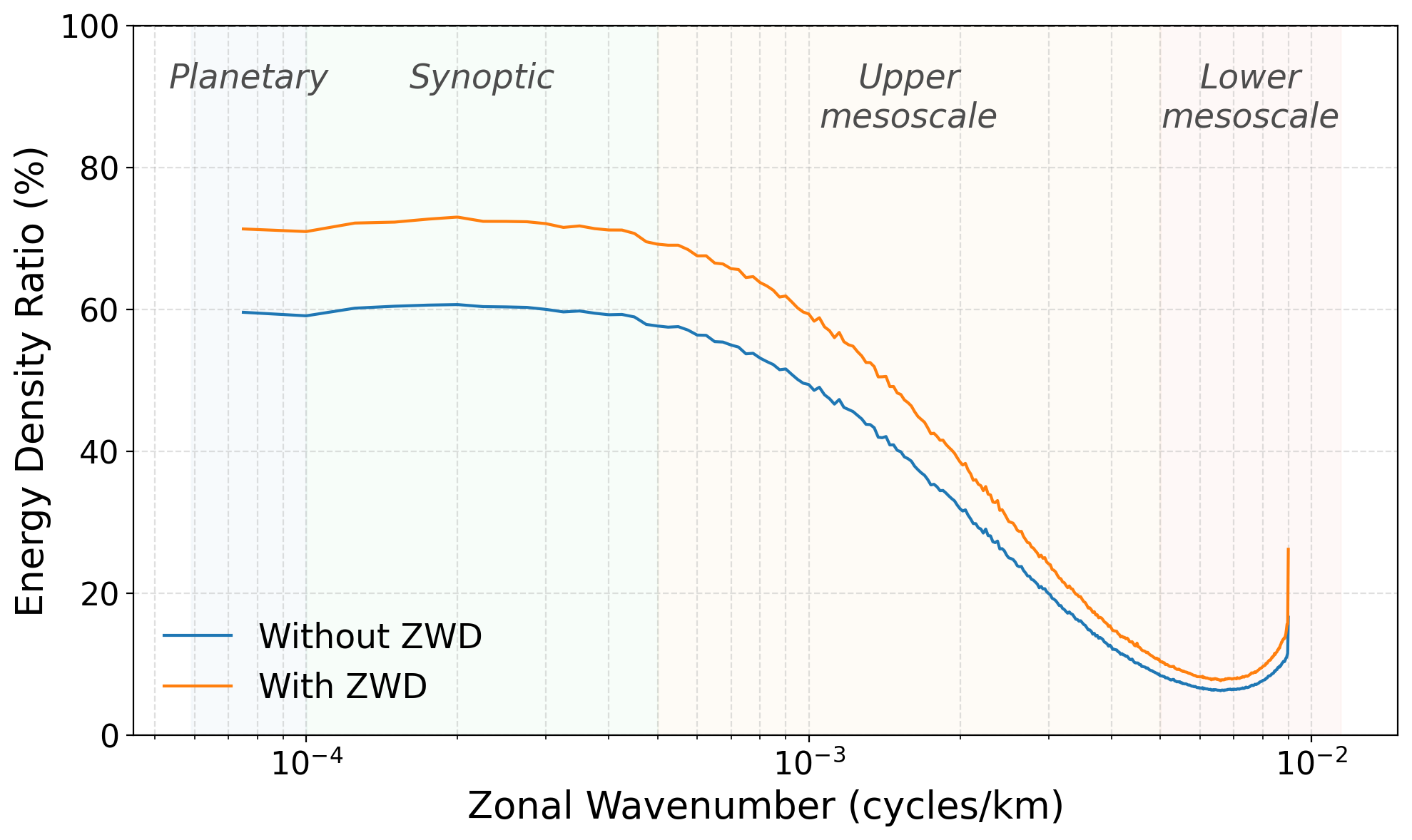}
\caption{\textbf{Including ZWD moves the predicted precipitation spectrum closer to the target at every scale.} Ratio of predicted to target latitude-weighted zonal power spectrum by wavenumber band for the main-analysis checkpoint ($100\%$ $=$ perfect match), for Model A (``With ZWD'') and Model B (``Without ZWD''). The corresponding checkpoint-averaged Log Spectral Distances are given in Supplementary Note~5.}
\label{fig:precip_spec}
\end{figure}

Including ZWD improves the predicted precipitation power spectrum on every spatial scale. The spectrum ratio (Figure~\ref{fig:precip_spec}) moves closer to unity for the ZWD model throughout the range, and the Log Spectral Distance decreases in every wavenumber band, with a total reduction of $-6.2\%$; the reduction is significant in every band under the checkpoint-wise paired $t$-test ($p \leq 0.023$; planetary and synoptic $p \leq 0.001$), with 8-9 of the ten checkpoints favouring ZWD throughout. The absolute LSD reduction is broadly comparable across bands ($\approx$0.06-0.08), so although the relative gain is highest at planetary ($-39.9\%$) and synoptic ($-33.3\%$) wavelengths, this reflects a much smaller baseline error at those scales rather than a benefit confined to them. The model is intrinsically blurrier on the mesoscale, and ZWD corrects the spectrum there by a similar absolute amount. See the full band-wise LSD values in Supplementary Note~5.

The improvement is therefore not limited to extreme-event intensity: ZWD also sharpens the spatial structure of the precipitation field, counteracting the over-smoothing typical of deterministic forecasts across the full range of scales. Together with the FSS improvement above, this indicates that ZWD's added information is largely structural rather than amplitude-correcting - improving where and at what scale precipitation is organized.

\subsection{The precipitation gain costs little and grows as model capacity shrinks}
\label{sec:robustness}

The precipitation gain is paid for by a small degradation of the variables Aurora already predicted well. Averaged over the same ten checkpoints, only the two 10\,m wind components degrade significantly on MAE, and by only ${\sim}1\%$; 2\,m temperature, mean sea-level pressure and specific humidity all fall within checkpoint-to-checkpoint noise. The more large-error-sensitive RMSE tells the same story but additionally flags 2\,m temperature as a small ($+3.9\%$) significant degradation. All five variables retain Pearson $R \geq 0.997$. Set against a precipitation improvement that is significant at every threshold, the trade-off is strongly asymmetric in favour of including ZWD. See the per-variable RMSE and MAE changes, with their paired-test $p$-values, in Supplementary Note~5.

The gain is also insensitive to how strongly ZWD is weighted in the training objective. A sweep over $\lambda_{\mathrm{ZWD}} \in \{1, 2, 3, 10\}$ shows that $\lambda_{\mathrm{ZWD}} \in \{1,2,3\}$ agree to within ${\sim}1\%$ on every precipitation metric and all outperform the baseline, with $\lambda_{\mathrm{ZWD}} = 2$ marginally best; only at $\lambda_{\mathrm{ZWD}} = 10$ does skill fall back to near-baseline, and the 75th-percentile ETS drops below it (see the full sweep in Supplementary Note~2). The value $\lambda_{\mathrm{ZWD}} = 2$ used in the main analysis is therefore a plateau rather than a tuned optimum.

Finally, the benefit of ZWD scales inversely with model capacity (Table~\ref{tab:crossscale}). The main analysis above uses the large Aurora variant ($\sim$1.3\,B parameters); repeating the With-versus-Without-ZWD comparison on the small variant ($\sim$110\,M parameters) shows that model scale dominates in absolute terms: the large baseline outperforms the small model with ZWD on every metric. But ZWD's \textit{relative} contribution is several times larger at the small scale on every metric, and the contrast is starkest in the extreme tail, where the 99th-percentile ETS gain reaches $+50.1\%$ against $+9.3\%$ in the large model. A third, precipitation-free ``surface-only'' 110\,M configuration further separates the cost of introducing the precipitation task from the cost of adding ZWD on top of it: the latter is essentially nil at this scale, with ZWD leaving the wind components unchanged and improving t2m and MSLP relative to the precipitation-only model (see further ablation studies in Supplementary Note~3). The 8.8\% headline gain reported above is thus a conservative figure for the capacity regime in which most operational and regional machine learning weather models currently sit.

\begin{table}[!ht]
\centering
\caption{\textbf{Effect of ZWD across model scales.} Headline precipitation metrics for the 110\,M and 1.3\,B Aurora variants, with and without ZWD. To keep the two scales comparable, all values are evaluated at the single main-analysis checkpoint; the 1.3\,B figures therefore differ slightly from the ten-checkpoint means of Tables~\ref{tab:precip_det} and~\ref{tab:precip_ets} (e.g.\ ETS$_{99}$ $+9.3\%$ here versus $+8.8\%$ checkpoint-averaged). The cross-scale ranking is unaffected. See additional metrics in Supplementary Notes~3 and~4.}
\label{tab:crossscale}
\small
\resizebox{\textwidth}{!}{%
\begin{tabular}{lccc|ccc}
\toprule
& \multicolumn{3}{c|}{110\,M Aurora} & \multicolumn{3}{c}{1.3\,B Aurora} \\
\cmidrule(lr){2-4}\cmidrule(lr){5-7}
Metric & Without ZWD & With ZWD & Rel.\ $\Delta$ & Without ZWD & With ZWD & Rel.\ $\Delta$ \\
\midrule
MAE [mm] $\downarrow$      & 0.265 & \textbf{0.247} & $-6.5\%$  & 0.192 & \textbf{0.189} & $-1.6\%$ \\
Normalised MSE $\downarrow$ & 0.445 & \textbf{0.417} & $-6.5\%$  & 0.260 & \textbf{0.248} & $-4.6\%$ \\
FSS 95\% $\uparrow$         & 0.735 & \textbf{0.843} & $+14.7\%$ & 0.915 & \textbf{0.931} & $+1.7\%$ \\
ETS 99\% $\uparrow$         & 0.203 & \textbf{0.304} & $+50.1\%$ & 0.386 & \textbf{0.422} & $+9.3\%$ \\
\bottomrule
\end{tabular}
}
\end{table}

\section{Discussion}

Adding a GNSS-derived observable to a weather foundation model improves precipitation forecasting, and the improvement is largest exactly where these models are weakest. ZWD is learned by Aurora at a skill level on par with already-pretrained variables, and including it during fine-tuning for 6-hour precipitation systematically improves precipitation forecasts, with the largest gains for the most extreme events (an 8.8\% increase in Equitable Threat Score at the 99th percentile) and a substantially more realistic precipitation power spectrum at planetary and synoptic scales. These gains are not only global: zooming in on three distinct high-impact events, the same benefit holds locally, with the error reduction concentrated in the heaviest-rain tail (Figure~\ref{fig:precip_regional}).

A plausible reason is that ZWD provides a direct, observation-derived column-moisture signal that the model would otherwise have to reconstruct from multi-level humidity, temperature, and pressure fields - a particular advantage for precipitation, which is itself a new downstream task for Aurora. This also reconciles the two improvements. At Aurora's 0.25$^{\circ}$ resolution, even the most extreme grid-cell rainfall is dominated by organized synoptic-scale systems (atmospheric rivers, fronts, tropical waves) whose moisture supply ZWD directly constrains. Improving the large-scale spectrum and sharpening extreme-event detection are therefore two sides of the same constraint. Consistent with this reading, the benefit grows as capacity shrinks (Table~\ref{tab:crossscale}): a smaller network has less representational budget to spend on reconstructing the column integral internally, so being handed it directly is worth more.

The present implementation relies on ZWDX, a gridded ZWD product that partially depends on ERA5 inputs and smooths individual station variability. Because its labels nonetheless carry the GNSS station-resolved moisture structure that ERA5 partly filters out (see Methods), the gains obtained despite this residual ERA5 dependence are a conservative lower bound on what raw, independent station data could provide. The natural next step is to feed GNSS station observations directly into the model (architecturally supported by the variable-specific ESFM encoder\citep{ozdemir2026esfm}), decoupling the GNSS signal from reanalysis and letting the model exploit the dense coverage achievable with next-generation low-cost receiver networks\citep{aichinger2023mpg, marut2022, soja2023igarss}. Further directions include physics-constrained training that enforces consistency between predicted ZWD and humidity column profiles, deeper case studies of the moisture build-up preceding high-impact events, operating at MSWEP's native $0.1^{\circ}$ resolution with a model able to ingest heterogeneous resolutions, and additional GNSS observables such as horizontal tropospheric gradients; the sub-hourly GNSS update rate ($<$5\,min) also matches convective time scales, motivating nowcasting.

More broadly, these findings show that variables not included in reanalysis products can carry physical information important enough to improve weather foundation models on the most challenging forecast tasks. ZWD is the example here, but the same principle would apply to other complementary observations such as radar reflectivity or dense environmental sensor networks. Systematic exploitation of these observational streams is a promising frontier for next-generation atmospheric machine learning.

\section{Methods}

\subsection{Aurora weather foundation model}

Aurora\citep{Bodnar2025} is a deep learning weather foundation model pretrained on ERA5\citep{hersbach2020}, two climate models from the Coupled Model Intercomparison Project Phase 6 (CMIP6)\citep{eyring2016} and operational forecasts. Given the atmospheric state at times $t-\Delta t$ and $t$, it predicts the state at $t+\Delta t$ on a $0.25^{\circ}$ global grid at $\Delta t = 6$-hour intervals. Surface and pressure-level variables are encoded into a shared latent representation, so new variables can be added by fine-tuning the model with an extended set of input and output variables. In every experiment reported here the Perceiver-based encoder, 3D Swin Transformer U-Net backbone and decoder are inherited from the pretrained checkpoint; the only modification is the extension of the linear surface encoder and decoder to accommodate the additional input and output channels, as shown in Figure~\ref{fig:overview} for the With-ZWD precipitation model. The Step~1 and baseline configurations differ from it only in which of the two new surface variables is present.

\subsection{ZWDX global gridded ZWD product}

To obtain the point-wise GNSS observations on a dense grid required by Aurora, we used ZWDX\citep{crocetti2024, crocetti2024a}, an XGBoost model trained on over 19{,}000 stations from the Nevada Geodetic Laboratory\citep{blewitt2018}. It learns the mapping from ERA5 specific humidity, location and time to GNSS-derived ZWD. The predictor inherits the spatio-temporal fidelity of the underlying station observations, while being able to provide ZWD values at any arbitrary point, including a regular grid. Because ZWDX is XGBoost-predicted from ERA5 inputs and station coordinates, GNSS information enters Aurora through the training-target labels rather than as raw, independent observations; the implications of this are discussed in Discussion. We exploit the ZWDX product on a $0.25^{\circ}$ grid (identical to ERA5) at 6-hour intervals to obtain spatially complete ZWD fields aligned with Aurora's pretraining inputs. Working from a gridded product, rather than raw station data, isolates the value of the GNSS ZWD from any additional gains attributable to a station-wise-input architecture.

\subsection{Precipitation target and regridding}

6-hour accumulated total precipitation is sourced from MSWEP V2\citep{beck2019}, a globally merged dataset combining reanalysis, satellite retrievals, and gauge observations at $0.1^{\circ}$ spatial resolution. MSWEP is preferred over ERA5 precipitation because ERA5 exhibits well-documented systematic biases in precipitation intensity and regional distribution\citep{beck2019, lavers2022}. Total precipitation was not included in Aurora's pretraining; predicting it therefore constitutes a genuinely new downstream task for the model, providing a demanding and realistic test of the value of ZWD as auxiliary information. Precipitation is log-transformed as $\log(1+x/\epsilon)$ with $\epsilon = 10^{-5}$, following \citet{rasp2024}.

The product is regridded to Aurora's $0.25^{\circ}$ grid by bilinear interpolation, so that each $0.25^{\circ}$ target cell aggregates approximately $(0.25/0.1)^2 \approx 6$ native cells. We use the same regridded MSWEP product as \citet{lehmann2025decoders}, accumulated here to 6-hour totals (the sum of two consecutive three-hour fields). Because precipitation is a mass-conserved quantity, bilinear interpolation does not exactly preserve the domain water budget: conservative (area-weighted) remapping is the mass-preserving alternative, and it additionally tends to attenuate the extreme upper tail of the distribution, because the value assigned to each coarse cell is an area average over its native cells rather than a point sample.

We nevertheless retain bilinear interpolation. Because Model A (with ZWD) and Model B (baseline) are trained and evaluated against the \textit{same} bilinearly regridded MSWEP target, any departure of that target from the ``true'' $0.25^{\circ}$ precipitation is common to both models and cancels in the with-versus-without-ZWD contrast that all reported gains measure; the regridding acts identically on both models' targets. We have not retrained under a conservative scheme, so the absolute scores and the precise magnitude of the ZWD gains may differ under a different regridding, whereas the with-versus-without-ZWD comparison does not.

\subsection{Experimental design}

\begin{figure}[!ht]
\centering
\includegraphics[width=0.9\linewidth]{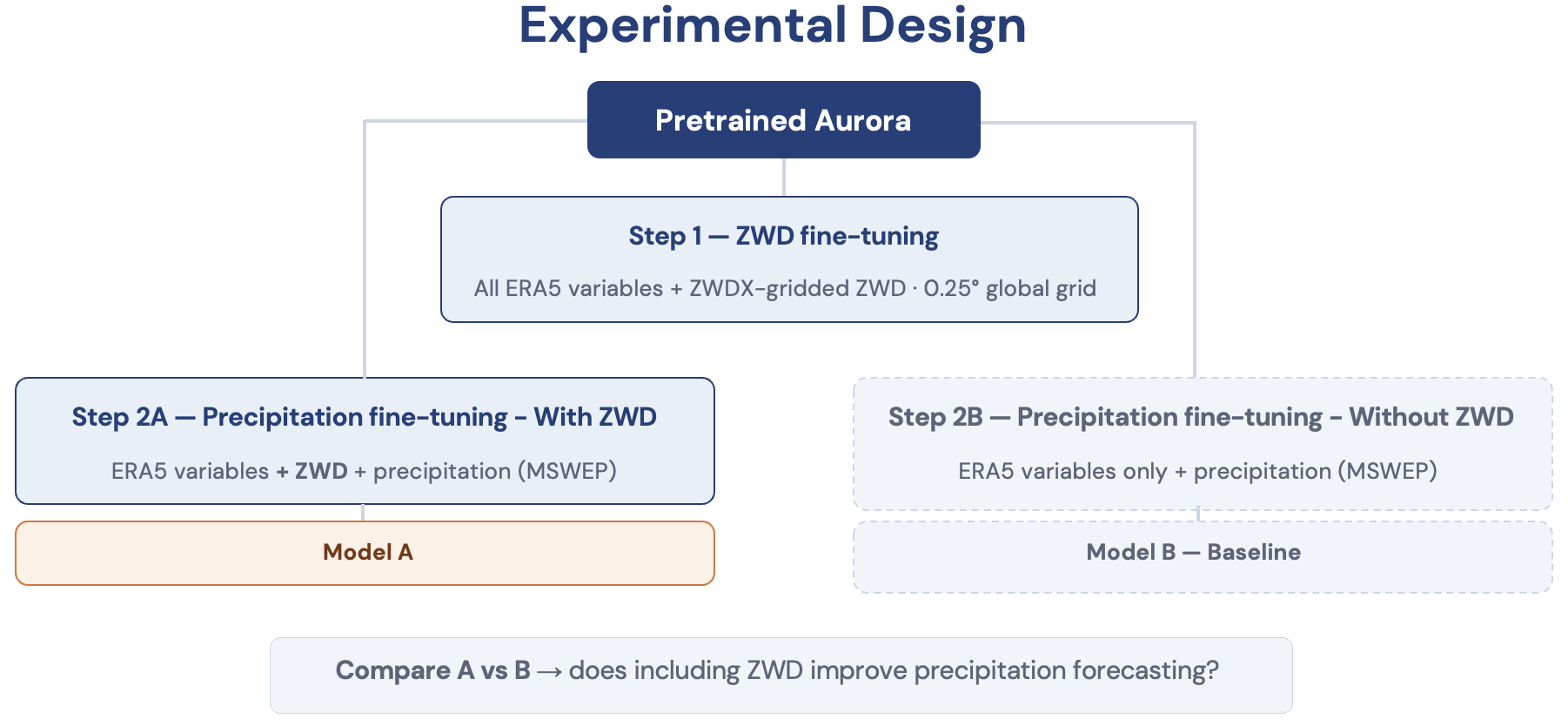}
\caption{\textbf{Schematic of the two-step fine-tuning workflow.} Starting from the pretrained Aurora checkpoint, Step~1 fine-tunes the model on the full ERA5 state augmented with ZWDX-gridded ZWD. Step~2 trains two precipitation models - Model~A (with ZWD) and Model~B (baseline) - that share optimiser, learning-rate schedule, number of steps and data, and differ only in whether ZWD is included as an additional input and auxiliary target.}
\label{fig:exp_design}
\end{figure}

The fine-tuning protocol comprises two stages (Figure~\ref{fig:exp_design}). Step~1 is a stand-alone experiment in which pretrained Aurora is fine-tuned on the full ERA5 surface and pressure-level state augmented with ZWDX-gridded ZWD as a new surface variable, establishing that the architecture can learn ZWD at a skill comparable to its standard pretrained variables. Step~2 consists of two parallel precipitation fine-tuning experiments. \textit{Model A} (``With ZWD'') and \textit{Model B} (baseline, ``Without ZWD'') are both initialised from the original Aurora pretrained checkpoint and trained with identical optimiser, learning-rate scheduler, number of training steps and data; they differ only in their variable set. In Model A, ZWD and precipitation are added as additional inputs and outputs, while only precipitation is used in Model B (both as input and output). The contrast between them therefore isolates the contribution of GNSS-derived ZWD to precipitation forecast skill.

The training objective is a per-variable, latitude-weighted Mean Absolute Error (MAE),
\begin{equation}
\mathcal{L} = \sum_{v} \lambda_v \, \frac{1}{N} \sum_{i} w(\phi_i)\,\mathrm{MAE}(\hat{x}_v, x_v),
\quad w(\phi) = \frac{\cos\phi}{\langle \cos\phi \rangle},
\label{eq:loss}
\end{equation}
where $\hat{x}_v$ and $x_v$ are the predicted and target fields for variable $v$ on the $0.25^{\circ}$ global grid, $\lambda_v$ is a per-variable loss weight, $\phi_i$ is the latitude of grid cell $i$, and $w(\phi)$ accounts for the decreasing area of grid cells towards the poles. The standard pretrained variables retain their pretraining weights, MSWEP precipitation enters with weight $1$, and ZWD enters with a tunable weight $\lambda_{\mathrm{ZWD}}$ set to $2$ in the main analysis (see the sensitivity analysis in Supplementary Note~2).

We use the Aurora large variant ($\sim$1.3\,B parameters) for the main analysis and the small variant ($\sim$110\,M parameters) for the cross-scale comparison. An alternative sequential protocol (initialising Model A from the Step~1 ZWD-augmented checkpoint) was also tested in preliminary experiments and yielded no additional skill gain. The 110\,M ablation additionally includes a third \textit{surface-only} configuration trained on the standard surface variables alone, with neither precipitation nor ZWD, which separates the cost of introducing the precipitation task itself from the additional cost of adding ZWD (Supplementary Note~3).

Training data span 2010-2018, with the lower bound set by the start of ZWDX availability; January 2019-March 2020 is used for validation, and April-December 2020 is held out for testing. All results are reported on the test set.

\subsection{Training configuration}

All fine-tuning experiments use the same optimiser and warmup schedule as Aurora's pretraining\citep{Bodnar2025} and optimise the per-variable weighted MAE objective of Equation~\eqref{eq:loss}.

For the 1.3\,B Aurora model, Step~1 (ZWD) uses a peak learning rate of $10^{-4}$, $4{,}000$ steps and $32$ GH200 GPUs (8 nodes $\times$ 4 GPUs) on the Alps supercomputer at the Swiss National Supercomputing Centre (CSCS); Step~2 (precipitation, Models A and B) uses the same learning rate, ${\sim}7{,}000$ steps and $16$ GPUs (4 nodes $\times$ 4 GPUs) for a 12-hour wall-time budget, with $\lambda_{\mathrm{ZWD}} = 2$ and precipitation weight $1$.

For the 110\,M small-model experiments, early-training instabilities required a smaller peak learning rate ($5 \times 10^{-5}$) and substantially smaller new-variable loss weights ($\lambda_{\mathrm{ZWD}} = \lambda_{\mathrm{precip}} = 0.1$); each experiment runs for 6 wall-clock hours on $16$ GPUs (4 nodes $\times$ 4 GPUs) for ${\sim}5{,}000$ steps. See the training curves for both steps in Supplementary Note~1.

\subsection{Evaluation metrics}
\label{sec:metrics}

Model performance is assessed using several complementary metrics. The \textit{Mean Absolute Error} (MAE) measures average pointwise prediction error in physical units (mm for ZWD and precipitation), and the \textit{Root Mean Squared Error} (RMSE) and \textit{Mean Squared Error} (MSE) further emphasise large errors. The \textit{Pearson correlation coefficient} ($R$) quantifies linear agreement between predicted and target fields. The \textit{Normalised MAE} expresses the physical MAE divided by the variable's standard deviation across the dataset, enabling comparison across variables with different physical scales, while the \textit{Relative MAE} is the MAE divided by the mean absolute value of the observations; an analogous normalised MSE is also reported.

The \textit{Fractions Skill Score} (FSS)\citep{roberts2008} is a neighbourhood-based metric that evaluates spatial structure by comparing the fraction of area exceeding a threshold between forecast and observation within windows of increasing size; we report FSS evaluated at the 95th percentile of the observed distribution, using a square neighbourhood window of side $9 \times 9$ grid points (corresponding to ${\sim}2.25^{\circ}$ at $0.25^{\circ}$ resolution). The \textit{Equitable Threat Score} (ETS)\citep{schaefer1990} is a categorical metric that measures the fraction of correctly forecast events above a given threshold while adjusting for hits expected by chance; ETS is reported at the 75th, 90th, 95th and 99th percentiles of the climatological distribution of the variable being verified - the precipitation distribution throughout the Step~2 results, and each variable's own observed distribution in the Step~1 comparison of Table~\ref{tab:zwd_skill} - and is particularly informative for severe-event verification.

\textit{Energy spectra} of precipitation are computed as the latitude-weighted (cosine of latitude) zonal power spectrum, obtained by applying a real-valued FFT along the longitude dimension and averaging $|\hat{f}(k)|^2$ across latitudes, to characterise the distribution of spatial variance across scales. Deviations of the predicted spectrum from the reference are summarised by the \textit{Log Spectral Distance} (LSD)\citep{gray1976, lam2023}, defined as the root-mean-square difference between the base-10 logarithms of the predicted and target power spectra,
\begin{equation}
\mathrm{LSD} = \sqrt{\overline{\left(\log_{10} S_{\mathrm{pred}} - \log_{10} S_{\mathrm{target}}\right)^2}},
\end{equation}
averaged over wavenumber within each of four bands: planetary ($5{,}000$--$20{,}000$\,km), synoptic ($1{,}000$--$5{,}000$\,km), upper mesoscale ($250$--$1{,}000$\,km) and lower mesoscale ($55$--$250$\,km). Where a \textit{total} (integrated) LSD is quoted, it is obtained by the same average taken over the full resolved wavenumber range ($55$--$20{,}000$\,km) at once, and is therefore not the sum or the mean of the four band values.

Where reported, the relative change is
\begin{equation}
\text{Rel.}\ \Delta = \frac{\text{With ZWD} - \text{Without ZWD}}{\text{Without ZWD}} \times 100\%,
\end{equation}
computed from the unrounded checkpoint means.

\subsection{Checkpoint ensemble and paired significance testing}
\label{sec:stats}

Step~2 metrics are reported as means over ten training checkpoints saved on the late-training plateau, all at $\lambda_{\mathrm{ZWD}} = 2$ (the checkpoint used for the single-checkpoint maps and diagnostics is one of the ten). For every metric we report the mean $\pm$ s.d.\ (computed with $\mathrm{ddof}=1$) across the ten checkpoints, the s.d.\ quantifying the intrinsic checkpoint-to-checkpoint variability of each model along its training trajectory.

Sampling ten checkpoints from a single training run per model is a deliberate compromise between statistical rigour and computational cost. The ideal design for separating the ZWD effect from training stochasticity would repeat each Step~2 training from several independent random seeds; at the scale of the 1.3\,B model, ten full retrainings per configuration is prohibitively expensive. Sampling ten checkpoints along the late-training plateau of one run per model is a far cheaper surrogate that still probes the stochastic-optimisation variability of the trajectory, and it is on this within-trajectory variability, rather than on true run-to-run variability, that the robustness assessment is based.

Model A and Model B do not share weights: they are distinct networks. They are, however, trained under an identical protocol (same pretrained initialisation, optimiser, learning-rate schedule, step budget and training data; Figure~\ref{fig:exp_design}), differing only in the presence of ZWD. A checkpoint is saved from each run at the \emph{same} ten training steps, so checkpoint~$i$ of Model~A and checkpoint~$i$ of Model~B are matched by training step: they occupy the same position along an otherwise identical trajectory. This step-matched correspondence is what defines a pair, and it is why the with-versus-without-ZWD contrast is naturally paired - differencing the two models at each matched step cancels the variation common to both runs at that step (schedule position, learning rate, cumulative data seen, stochastic-optimisation phase) - and isolates the ZWD effect. We therefore report how many of the ten checkpoints favour the ZWD model together with the two-sided paired $t$-test $p$-value across checkpoints. This is the appropriate test for a per-checkpoint difference and is more sensitive than comparing the two means against their larger marginal spread; a paired $t$-test moreover remains valid under any pairing, so the residual divergence of the two trajectories once ZWD is added can only reduce its power, never inflate its false-positive rate.

The per-model checkpoint scatter is itself sizeable (e.g.\ $\pm0.05$ on ETS$_{99}$), comparable to the mean With-minus-Without gap, so the two models' $\pm$s.d.\ ranges overlap; the gain is nonetheless robust precisely because it is systematic within each matched pair, which is what the paired test captures. The With-ZWD model is moreover the more stable of the two, with a smaller s.d.\ on every precipitation metric. See the full checkpoint-ensemble tables in Supplementary Note~5.

\section*{Data availability}
The global ZWDX product used in this study is described in \citet{crocetti2024} and is available at \url{https://doi.org/10.1186/s40623-024-02104-6}. MSWEP V2 precipitation data are described in \citet{beck2019} and are available at \url{http://www.gloh2o.org/mswep/}. ERA5 reanalysis data are available from the Copernicus Climate Data Store (\url{https://cds.climate.copernicus.eu}). The weights of the pretrained Aurora model are available from the Microsoft Aurora repository (\url{https://github.com/microsoft/aurora}). The trained model checkpoints generated in this study are available from the corresponding author upon reasonable request.

\section*{Code availability}
The fine-tuning, evaluation and figure-generation code for this study is available at \url{https://github.com/swiss-ai/zwd-into-aurora}. It builds on the pretrained model and inference code of the Microsoft Aurora repository (\url{https://github.com/microsoft/aurora}) and on the Earth System Foundation Model (ESFM) training framework (\url{https://github.com/swiss-ai/ESFM}).

\section*{Acknowledgements}
This work was supported under project IDs a122 and a0196 as part of the Swiss AI Initiative, through a grant from the ETH Domain and computational resources provided by the Swiss National Supercomputing Centre (CSCS) under the Alps infrastructure. Contributions of L.T. were supported by the Swiss National Science Foundation (SNSF) under grant number 225851. Contributions of F.L. were primarily supported by the ETH AI Center through their ETH AI Center postdoctoral fellowship. The authors thank Firat Ozdemir and Yun Cheng for their help in setting up the experiments, and Simon Adamov for his assistance with the evaluation. The authors used Claude (Anthropic) for language editing and code debugging during manuscript preparation; the authors reviewed all output and take full responsibility for the content.

\section*{Author contributions}
L.T. conceived the study, developed the methodology and software, performed the experiments, formal analysis and validation, produced the figures, and wrote the original draft. F.L. contributed to the methodology and software, supported validation, supervised the work, and reviewed and edited the manuscript. L.C. provided the ZWDX gridded ZWD product (resources and data curation) and reviewed and edited the manuscript. B.S. conceived the study jointly with L.T., supervised the project, administered it, acquired funding, provided resources, and reviewed and edited the manuscript. All authors read and approved the final manuscript.

\section*{Competing interests}
The authors declare no competing interests.

\bibliographystyle{unsrtnat}
\bibliography{references}

\clearpage

\setcounter{figure}{0}
\setcounter{table}{0}
\renewcommand{\thefigure}{S\arabic{figure}}
\renewcommand{\thetable}{S\arabic{table}}
\renewcommand{\figurename}{Supplementary Fig.}
\renewcommand{\tablename}{Supplementary Table}
\linespread{1.15}\selectfont

\begin{center}
{\bfseries\large Supplementary Information}\\[0.4em]
{\itshape Integrating GNSS-Derived Zenith Wet Delay into a Weather Foundation Model Improves Precipitation Forecasting}
\end{center}

\vspace{1em}

\noindent\textbf{Contents}

\begin{itemize}\setlength{\itemsep}{1pt}
\item Supplementary Note 1 - Training curves
\item Supplementary Note 2 - Loss-weight sensitivity (1.3\,B model)
\item Supplementary Note 3 - Three-way small-model ablation (110\,M): surface-only vs Without ZWD vs With ZWD
\item Supplementary Note 4 - Cross-scale comparison (110\,M vs 1.3\,B parameters)
\item Supplementary Note 5 - Checkpoint variability of the ZWD effect, spectral skill and cost to the pretrained variables (1.3\,B model, $\lambda_{\mathrm{ZWD}} = 2$)
\item Supplementary Note 6 - Rollout skill as a function of lead time
\item Supplementary Note 7 - Temporal consistency and learning-quality examples
\item Supplementary Note 8 - Regional extreme-event case studies
\item Supplementary Figures S1-S15; Supplementary Tables S1-S15
\end{itemize}

\noindent This Supplementary Information collects sensitivity analyses, robustness checks, extended result tables and qualitative examples accompanying the main article. It contains supporting detail only: every result on which a claim in the article rests, and all descriptions of Methods, are given in the main manuscript above. References of the form ``main text'' point to the main article of this document.

\clearpage

\suppnote{Training curves}
\label{app:training}

\begin{figure}[htbp]
\centering
\begin{subfigure}{0.92\linewidth}
\centering
\includegraphics[width=\linewidth]{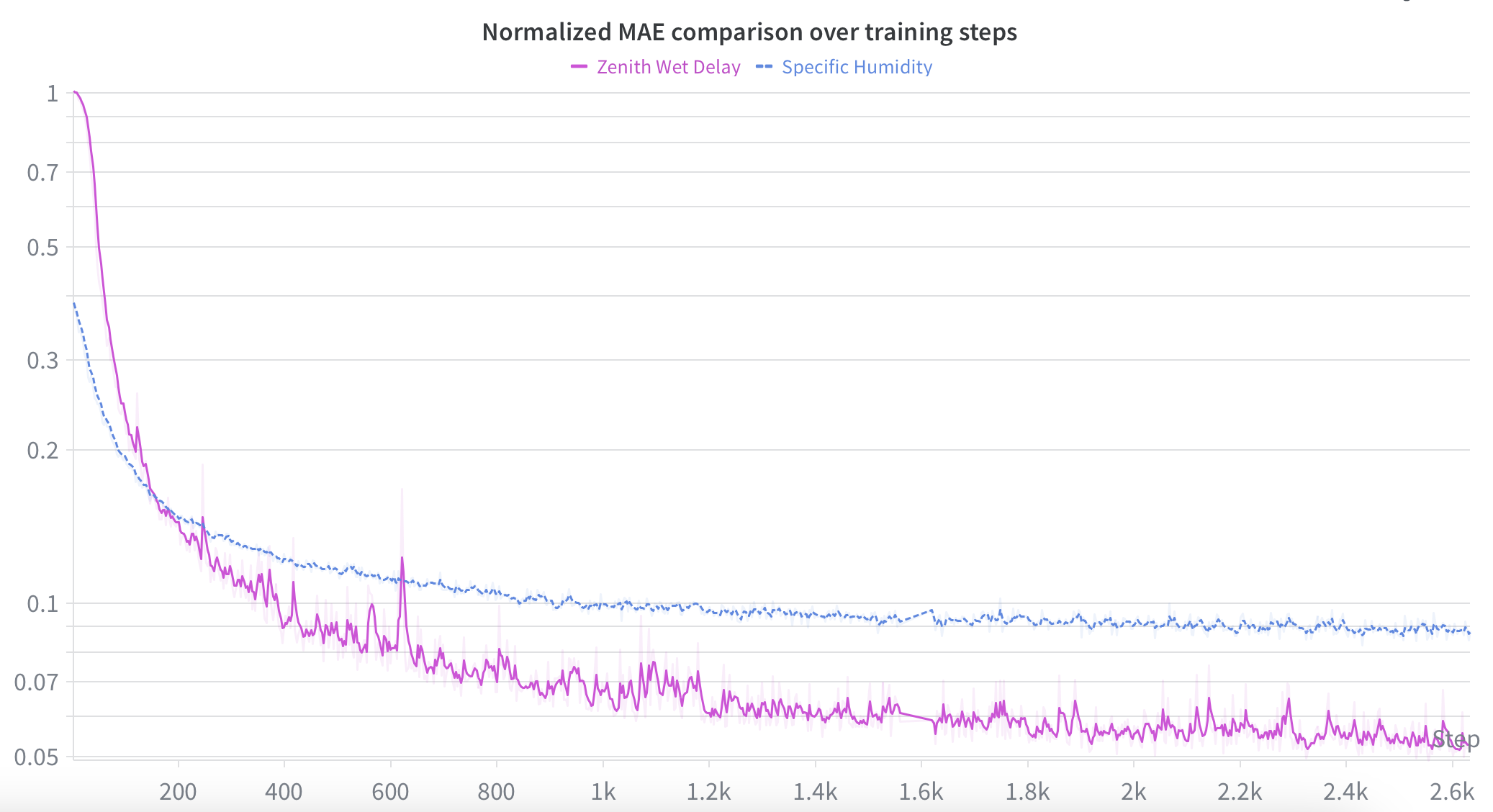}
\caption{Step~1: ZWD (magenta) and specific humidity (blue).}
\label{fig:training_curve_zwd}
\end{subfigure}

\vspace{0.8em}

\begin{subfigure}{0.92\linewidth}
\centering
\includegraphics[width=\linewidth]{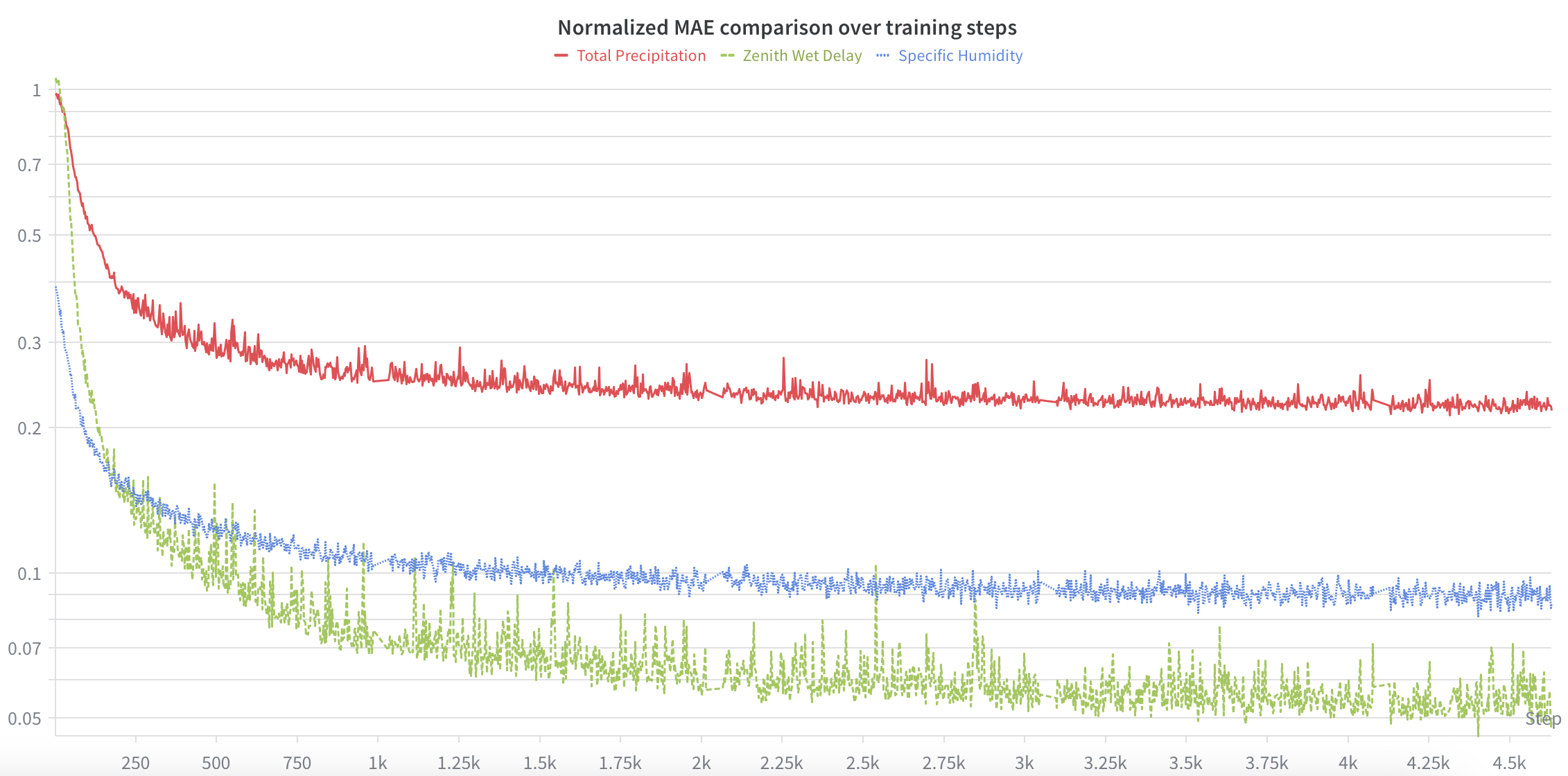}
\caption{Step~2 (Model~A): total precipitation (red), ZWD (green) and specific humidity (blue).}
\label{fig:training_curve_precip}
\end{subfigure}
\caption{\textbf{Normalised MAE over training steps for the 1.3\,B Aurora} (log scale; lower is better). \textbf{a} Step~1 ZWD fine-tuning; \textbf{b} Step~2 precipitation fine-tuning with ZWD. The horizontal axis is truncated to the region where the curves vary; training continued to the full step counts given in Methods, over which the normalised MAE remained essentially flat.}
\label{fig:training_curves}
\end{figure}

Supplementary Fig.~\ref{fig:training_curves} shows the evolution of the normalised MAE over training steps for the 1.3\,B model; the corresponding learning rates, step counts and hardware are given in Methods, ``Training configuration''. In Step~1 (Supplementary Fig.~\ref{fig:training_curve_zwd}), ZWD converges within the first few hundred steps to a normalised MAE comparable to that of specific humidity, the closest pretrained moisture analogue. In Step~2 (Supplementary Fig.~\ref{fig:training_curve_precip}), total precipitation (a genuinely new task) converges to a substantially higher normalised MAE than the auxiliary ZWD and specific humidity targets, reflecting the intrinsic difficulty of precipitation, while ZWD and humidity remain at the low error levels established during Step~1.

\suppnote{Loss-weight sensitivity (1.3\,B model)}
\label{app:lw}

In the Step~2 precipitation fine-tuning experiment, Model A predicts both 6-hour accumulated precipitation and ZWD, the latter acting as an auxiliary target whose contribution to the joint training objective is set by the loss weight $\lambda_{\mathrm{ZWD}}$. We compare four values ($\lambda_{\mathrm{ZWD}} \in \{1, 2, 3, 10\}$) against the baseline Model B (no ZWD) on the 1.3\,B Aurora; $\lambda_{\mathrm{ZWD}} = 2$ is the value adopted in the main analysis. All values in this note are evaluated at the single main-analysis checkpoint, whereas the headline metrics in the main text and in Supplementary Note~5 are averaged over ten training checkpoints; the baseline and $\lambda_{\mathrm{ZWD}} = 2$ figures here therefore differ slightly from those ten-checkpoint means.

Supplementary Table~\ref{tab:app_lw_det} summarises deterministic precipitation skill: $\lambda_{\mathrm{ZWD}} = 2$ achieves the best score on every metric except Pearson $R$ (where the baseline leads marginally), and $\lambda_{\mathrm{ZWD}} = 10$ degrades to near-baseline or worse. Supplementary Table~\ref{tab:app_lw_ets} confirms the same picture for ETS at every threshold; at the 75th percentile $\lambda_{\mathrm{ZWD}} = 10$ drops below baseline. Spectral fidelity (Supplementary Table~\ref{tab:app_lw_lsd}) is best at $\lambda_{\mathrm{ZWD}} = 1$ and 2, with the differences between them under 1\%. Finally, degradation of the pretrained variables (Supplementary Table~\ref{tab:app_lw_other}) is monotonic with the loss weight for the wind components and non-monotonic for t2m and MSLP; at $\lambda_{\mathrm{ZWD}} = 2$ the costs are moderate ($+1.4\%$ to $+7.7\%$ across the four variables). Across all of these criteria, $\lambda_{\mathrm{ZWD}} = 2$ is the optimal choice, and the plateau across $\{1,2,3\}$ shows the result is not sensitive to its precise value.

\begin{table}[h!]
\caption{Deterministic precipitation metrics for the 1.3\,B Aurora across ZWD loss weights (lw).}
\label{tab:app_lw_det}
\centering
\begin{tabular}{lccccc}
\toprule
Metric & No ZWD & lw=1 & lw=2 & lw=3 & lw=10 \\
\midrule
MAE [mm]               & 0.192 & 0.190 & \textbf{0.189} & 0.190 & 0.192 \\
RMSE [mm]              & 1.015 & 0.997 & \textbf{0.991} & 0.999 & 1.004 \\
Pearson $R$            & \textbf{0.853} & 0.851 & 0.852 & 0.851 & 0.848 \\
FSS 95\%               & 0.915 & 0.929 & \textbf{0.931} & 0.929 & 0.925 \\
Relative MAE           & 0.269 & 0.266 & \textbf{0.265} & 0.266 & 0.270 \\
Normalised MSE         & 0.260 & 0.250 & \textbf{0.248} & 0.252 & 0.254 \\
\bottomrule
\end{tabular}
\end{table}

\begin{table}[h!]
\caption{ETS for precipitation at four climatological percentiles (pct) across ZWD loss weights (lw).}
\label{tab:app_lw_ets}
\centering
\begin{tabular}{lccccc}
\toprule
Threshold & No ZWD & lw=1 & lw=2 & lw=3 & lw=10 \\
\midrule
75th pct & 0.695 & 0.699 & \textbf{0.703} & \textbf{0.703} & 0.692 \\
90th pct & 0.654 & 0.666 & \textbf{0.669} & 0.668 & 0.657 \\
95th pct & 0.574 & 0.598 & \textbf{0.600} & 0.598 & 0.587 \\
99th pct & 0.386 & 0.420 & \textbf{0.422} & 0.415 & 0.414 \\
\bottomrule
\end{tabular}
\end{table}

\begin{table}[h!]
\caption{Log Spectral Distance (LSD) for precipitation across ZWD loss weights (lw).}
\label{tab:app_lw_lsd}
\centering
\begin{tabular}{lccccc}
\toprule
Band & No ZWD & lw=1 & lw=2 & lw=3 & lw=10 \\
\midrule
Planetary       & 0.223 & 0.155 & \textbf{0.145} & 0.165 & 0.156 \\
Synoptic        & 0.259 & 0.187 & \textbf{0.182} & 0.203 & 0.189 \\
Upper mesoscale & 0.636 & 0.562 & \textbf{0.558} & 0.589 & 0.561 \\
Lower mesoscale & 1.076 & \textbf{0.992} & 1.001 & 1.048 & 1.013 \\
Total           & 0.986 & \textbf{0.906} & 0.914 & 0.957 & 0.924 \\
\bottomrule
\end{tabular}
\end{table}

\begin{table}[h!]
\caption{Normalised MAE of the pretrained surface variables and of precipitation across ZWD loss weights (lw).}
\label{tab:app_lw_other}
\centering
\begin{tabular}{lccccc}
\toprule
Variable & No ZWD & lw=1 & lw=2 & lw=3 & lw=10 \\
\midrule
u10    & \textbf{0.0363} & 0.0366 & 0.0368 & 0.0372 & 0.0383 \\
v10    & \textbf{0.0436} & 0.0440 & 0.0444 & 0.0444 & 0.0458 \\
t2m    & \textbf{0.0087} & 0.0092 & 0.0090 & 0.0090 & 0.0094 \\
MSLP   & \textbf{0.0091} & 0.0095 & 0.0098 & 0.0092 & 0.0098 \\
Precip & 0.0963 & 0.0952 & \textbf{0.0950} & 0.0953 & 0.0966 \\
\bottomrule
\end{tabular}
\end{table}

\suppnote{Three-way small-model ablation (110\,M): surface-only vs Without ZWD vs With ZWD}
\label{app:smallmodel}

We repeat the main With-versus-Without-ZWD comparison on the small (110\,M) Aurora variant, contrasting the ``Without ZWD'' model (precipitation, no ZWD) with the ``With ZWD'' model exactly as in the 1.3\,B analysis. The only new element here is a third configuration - a \textit{surface-only} model trained on the standard surface variables alone, with neither precipitation nor ZWD - which lets us separate the cost of introducing the precipitation task itself from the additional cost of adding ZWD on the pretrained variables, a distinction the two-model comparison cannot make on its own. Headline figures from this note are reported in Table~4 of the main text.

\subsection{Precipitation skill}

Adding ZWD to the precipitation model improves every deterministic precipitation metric, including Pearson $R$, which (unlike in the 1.3\,B case) does not decrease (Supplementary Table~\ref{tab:app_smallmodel_det}). The FSS improvement is particularly large ($+14.7\%$). The normalised MSE reduction ($-6.5\%$) exceeds that of the 1.3\,B model, a cross-scale contrast examined in Supplementary Note~4. ETS (Supplementary Table~\ref{tab:app_smallmodel_ets}) shows the same threshold-dependent pattern as in the large model, but with substantially larger relative gains at every threshold: at the 99th percentile ETS improves by $+50\%$, the largest relative gain at any threshold. The spectral improvements (Supplementary Table~\ref{tab:app_smallmodel_lsd}) are large: the ``Without ZWD'' model produces fields with substantial spectral distortion ($\mathrm{LSD} > 0.45$) at planetary and synoptic scales, and adding ZWD reduces this by roughly 60\% at those scales, with a total LSD reduction of 38.6\%.

\begin{table}[h!]
\caption{Deterministic precipitation metrics for the 110\,M Aurora: Without ZWD vs With ZWD.}
\label{tab:app_smallmodel_det}
\centering
\begin{tabular}{lccc}
\toprule
Metric & Without ZWD & With ZWD & Relative change \\
\midrule
MAE [mm]               & 0.265 & \textbf{0.247} & $-6.5\%$ \\
RMSE [mm]              & 1.329 & \textbf{1.285} & $-3.3\%$ \\
MSE [mm$^2$]           & 1.767 & \textbf{1.652} & $-6.5\%$ \\
Pearson $R$            & 0.719 & \textbf{0.728} & $+1.3\%$ \\
FSS 95\%               & 0.735 & \textbf{0.843} & $+14.7\%$ \\
Relative MAE           & 0.370 & \textbf{0.346} & $-6.5\%$ \\
Normalised MAE         & 0.133 & \textbf{0.124} & $-6.5\%$ \\
Normalised MSE         & 0.445 & \textbf{0.417} & $-6.5\%$ \\
\bottomrule
\end{tabular}
\end{table}

\begin{table}[h!]
\caption{ETS for precipitation at four climatological percentiles (pct) for the 110\,M Aurora.}
\label{tab:app_smallmodel_ets}
\centering
\begin{tabular}{lccc}
\toprule
Threshold & Without ZWD & With ZWD & Relative change \\
\midrule
75th pct & 0.527 & \textbf{0.587} & $+11.4\%$ \\
90th pct & 0.453 & \textbf{0.533} & $+17.7\%$ \\
95th pct & 0.354 & \textbf{0.453} & $+27.8\%$ \\
99th pct & 0.203 & \textbf{0.304} & $+50.1\%$ \\
\bottomrule
\end{tabular}
\end{table}

\begin{table}[h!]
\caption{Log Spectral Distance (LSD) for precipitation in the 110\,M Aurora. \textit{Total} is LSD integrated over the full wavenumber range, not the sum of band values.}
\label{tab:app_smallmodel_lsd}
\centering
\begin{tabular}{lccc}
\toprule
Band & Without ZWD & With ZWD & Relative change \\
\midrule
Planetary       & 0.457 & \textbf{0.185} & $-59.5\%$ \\
Synoptic        & 0.465 & \textbf{0.177} & $-61.9\%$ \\
Upper mesoscale & 0.752 & \textbf{0.418} & $-44.4\%$ \\
Lower mesoscale & 0.606 & \textbf{0.386} & $-36.4\%$ \\
\textbf{Total}  & 0.631 & \textbf{0.388} & $\mathbf{-38.6\%}$ \\
\bottomrule
\end{tabular}
\end{table}

\begin{figure}[htbp]
\centering
\includegraphics[width=\linewidth]{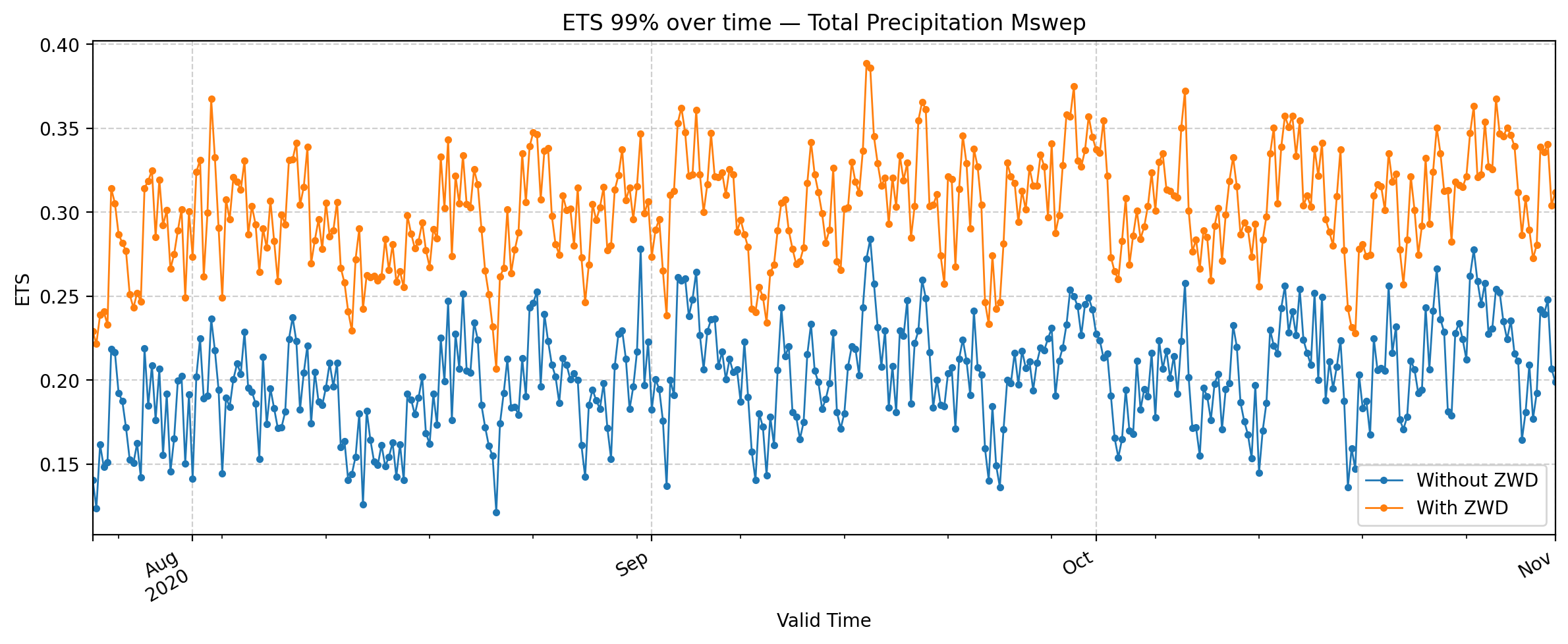}
\caption{\textbf{ETS at the 99th percentile over the test period for the 110\,M Aurora}, Without ZWD versus With ZWD.}
\label{fig:app_smallmodel_ets_ts}
\end{figure}

Supplementary Fig.~\ref{fig:app_smallmodel_ets_ts} shows ETS at the 99th percentile over the test period for the two 110\,M precipitation models. The ``With ZWD'' model outperforms the ``Without ZWD'' model at essentially every timestep, mirroring the behaviour of the 1.3\,B model (Supplementary Fig.~\ref{fig:app_ets99_ts}) but with a substantially larger margin.

\subsection{Cost to pretrained variables}

Adding precipitation and ZWD does not materially degrade the skill of the standard pretrained surface variables (Supplementary Table~\ref{tab:app_smallmodel_other}). Wind components remain at the surface-only skill level or slightly better; t2m shows a small increase in normalised MAE relative to surface-only ($+5.7\%$), substantially better than in the ``Without ZWD'' model; MSLP is comparable across all three configurations. Adding ZWD on top of precipitation therefore does not introduce meaningful additional degradation beyond that already caused by precipitation alone - in contrast to the 1.3\,B model, where a small but significant wind-component cost is observed (Supplementary Table~\ref{tab:app_precip_cost}).

\begin{table}[h!]
\caption{Normalised MAE of the pretrained surface variables across the three 110\,M configurations.}
\label{tab:app_smallmodel_other}
\centering
\begin{tabular}{lccc}
\toprule
Variable & Surface-only & Without ZWD & With ZWD \\
\midrule
u10   & 0.0488 & 0.0466 & \textbf{0.0464} \\
v10   & 0.0576 & 0.0567 & \textbf{0.0559} \\
t2m   & \textbf{0.0122} & 0.0161 & 0.0129 \\
MSLP  & 0.0140 & 0.0145 & \textbf{0.0124} \\
\bottomrule
\end{tabular}
\end{table}

\suppnote{Cross-scale comparison (110\,M vs 1.3\,B parameters)}
\label{app:crossscale}

We place the small (110\,M) and large (1.3\,B) Aurora variants side by side. The small-model entries are the ``Without ZWD'' and ``With ZWD'' configurations from the ablation in Supplementary Note~3; the large-model entries are Models~B and~A from the main analysis, evaluated here at the single main-analysis checkpoint for consistency with the small-model figures. The large-model values therefore differ slightly from the ten-checkpoint means reported in Tables~2 and~3 of the main text (e.g.\ the ETS$_{99}$ gain is $+9.3\%$ at this checkpoint versus $+8.8\%$ averaged over the ten checkpoints); the cross-scale ranking is unaffected. The headline version of this comparison is Table~4 of the main text; Supplementary Table~\ref{tab:app_crossscale_det} and Supplementary Table~\ref{tab:app_crossscale_ets99} give it in full.

Three patterns are evident. First, model scale dominates: the large baseline outperforms the small model with ZWD on every metric. Second, ZWD's \textit{relative} benefit is consistently larger at the small scale, across normalised MSE, FSS and threshold skill alike. Third, this scale contrast is starkest at the 99th-percentile ETS, where the gain reaches $+50\%$ in the small model against $+9.3\%$ in the large one.

\begin{table}[h!]
\caption{Headline deterministic precipitation metrics across model scales, with and without ZWD. The full metric set for the small model is given in Supplementary Table~\ref{tab:app_smallmodel_det}.}
\label{tab:app_crossscale_det}
\centering
\resizebox{\textwidth}{!}{%
\begin{tabular}{lcccc}
\toprule
Metric & Small (Without ZWD) & Small (With ZWD) & Large (Without ZWD) & Large (With ZWD) \\
\midrule
MAE [mm]        & 0.265 & 0.247 & 0.192 & \textbf{0.189} \\
FSS 95\%        & 0.735 & 0.843 & 0.915 & \textbf{0.931} \\
Normalised MSE  & 0.445 & 0.417 & 0.260 & \textbf{0.248} \\
\bottomrule
\end{tabular}
}
\end{table}

\begin{table}[h!]
\caption{Equitable Threat Score (ETS) at the 99th percentile, across model scales and ZWD configurations.}
\label{tab:app_crossscale_ets99}
\centering
\begin{tabular}{lcc}
\toprule
Configuration & ETS 99\% & Relative change \\
\midrule
Small (Without ZWD) & 0.203 & - \\
Small (With ZWD)    & 0.304 & $+50.1\%$ \\
Large (Without ZWD) & 0.386 & - \\
Large (With ZWD)    & 0.422 & $+9.3\%$ \\
\bottomrule
\end{tabular}
\end{table}

The per-timestep view in Supplementary Fig.~\ref{fig:app_crossscale_ts} confirms that these conclusions hold throughout the test period rather than on the test-set average alone. At essentially every verification time the four configurations remain cleanly ordered (large (With ZWD) $\geq$ large (Without ZWD) $>$ small (With ZWD) $>$ small (Without ZWD)) for both ETS at the 99th percentile (Supplementary Fig.~\ref{fig:app_crossscale_ets_ts}) and FSS at the 95th percentile (Supplementary Fig.~\ref{fig:app_crossscale_fss_ts}). The ZWD-induced gain is visibly largest for the small model, consistent with the inverse capacity scaling reported above.

\begin{figure}[htbp]
\centering
\begin{subfigure}{\linewidth}
\centering
\includegraphics[width=\linewidth]{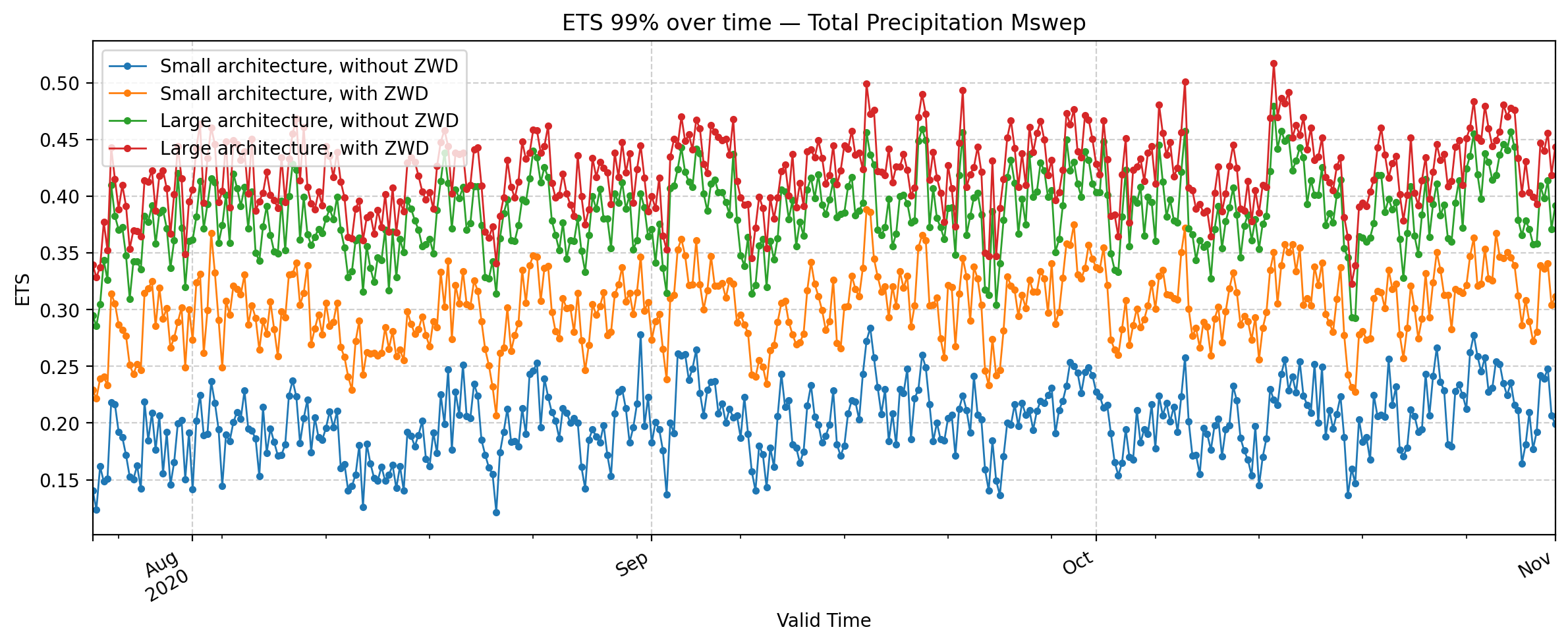}
\caption{ETS at the 99th percentile.}
\label{fig:app_crossscale_ets_ts}
\end{subfigure}

\vspace{0.8em}

\begin{subfigure}{\linewidth}
\centering
\includegraphics[width=\linewidth]{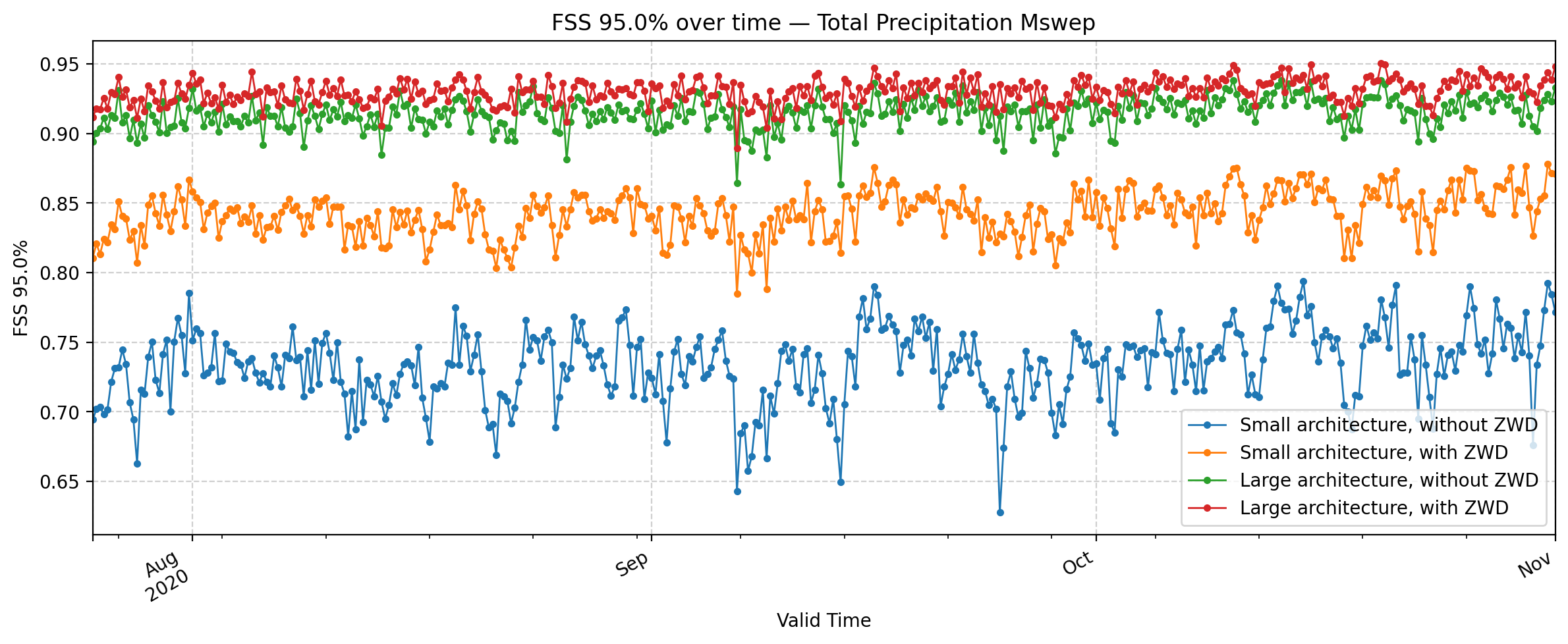}
\caption{FSS at the 95th percentile.}
\label{fig:app_crossscale_fss_ts}
\end{subfigure}
\caption{\textbf{Precipitation skill over the test period for the four model configurations} (small/large $\times$ with/without ZWD). The configuration ordering is preserved at essentially every timestep.}
\label{fig:app_crossscale_ts}
\end{figure}

\suppnote{Checkpoint variability of the ZWD effect, spectral skill and cost to the pretrained variables (1.3\,B model, $\lambda_{\mathrm{ZWD}} = 2$)}
\label{app:checkpoint}

The deterministic and threshold-based gains in the main text are reported as means over ten training checkpoints, all at $\lambda_{\mathrm{ZWD}} = 2$ and on the late-training plateau. The construction of this checkpoint ensemble, the rationale for the paired test and its validity are described in Methods, ``Checkpoint ensemble and paired significance testing''. This note documents the ensemble in full: the per-model spread and the paired significance of the With-versus-Without-ZWD difference, together with the checkpoint-averaged band-wise spectral skill and the cost to the pretrained variables underlying the summary statements in Results. This analysis complements the loss-weight sweep of Supplementary Note~2, which instead varies $\lambda_{\mathrm{ZWD}}$ at a fixed checkpoint: here $\lambda_{\mathrm{ZWD}}$ is held fixed and the checkpoint varies, so the two together bound the sensitivity of the result to both the hyperparameter and the training trajectory.

\subsection{Deterministic skill and paired significance}

Supplementary Table~\ref{tab:app_ckpt_var_det} reports the precipitation deterministic skill of each model as a mean $\pm$ s.d., and Supplementary Table~\ref{tab:app_ckpt_var_paired} the paired head-to-head comparison for RMSE; the corresponding ETS paired results (number of checkpoints favouring ZWD and paired $p$-value at each threshold) are given in Table~3 of the main text and are not repeated here. The precipitation improvement is reproduced at almost every checkpoint (8-9 of the ten for every metric) and is statistically significant under the paired test both for RMSE ($9/10$ checkpoints favour ZWD, $p = 0.032$) and at every ETS threshold ($p \leq 0.013$; main text).

\begin{table}[h!]
\caption{Precipitation deterministic skill as mean $\pm$ s.d.\ over ten training checkpoints (1.3\,B model, $\lambda_{\mathrm{ZWD}}=2$).}
\label{tab:app_ckpt_var_det}
\centering
\begin{tabular}{lcc}
\toprule
Metric & Without ZWD (Model B) & With ZWD (Model A) \\
\midrule
MAE [mm]     & $0.195 \pm 0.007$ & $\mathbf{0.193 \pm 0.004}$ \\
RMSE [mm]    & $1.015 \pm 0.034$ & $\mathbf{0.997 \pm 0.019}$ \\
MSE [mm$^2$] & $1.032 \pm 0.071$ & $\mathbf{0.994 \pm 0.038}$ \\
\bottomrule
\end{tabular}
\end{table}

\begin{table}[h!]
\caption{Paired checkpoint comparison of precipitation RMSE (1.3\,B model, $\lambda_{\mathrm{ZWD}}=2$): mean $\pm$ s.d.\ for each model, relative change, number of checkpoints favouring ZWD, and two-sided paired $t$-test $p$-value (lower RMSE is better). The corresponding ETS paired results are reported in Table~3 of the main text. $^{*}\,p<0.05$.}
\label{tab:app_ckpt_var_paired}
\centering
\begin{tabular}{lccccc}
\toprule
Metric & Without ZWD & With ZWD & Rel.\ $\Delta$ & ZWD wins & paired $p$ \\
\midrule
RMSE [mm]  & $1.015 \pm 0.034$ & $\mathbf{0.997 \pm 0.019}$ & $-1.8\%$ & 9/10 & $0.032^{*}$ \\
\bottomrule
\end{tabular}
\end{table}

\subsection{Spectral skill by wavenumber band}

The spectral improvement is checkpoint-robust in the same way (Supplementary Table~\ref{tab:app_precip_lsd}). The band-wise Log Spectral Distance decreases with ZWD at 8-9 of the ten checkpoints in every band, with a two-sided paired-test significance of $p \leq 0.023$ (planetary and synoptic $p \leq 0.001$). As for ETS, the per-checkpoint spread of the LSD is large relative to the mean With-minus-Without gap (especially at planetary and synoptic scales, where the baseline LSD is itself small), so the improvement is significant through its consistency across checkpoints rather than by separation of the marginal $\pm$s.d.\ ranges. The corresponding spectrum ratio is Fig.~6 of the main text.

\begin{table}[h!]
\centering
\caption{Step~2 spectral precipitation skill (Log Spectral Distance) on the held-out test set, as a mean $\pm$ s.d.\ over ten training checkpoints; bold marks the better model (lower LSD is a more faithful spectrum). The reduction is significant in every band under the checkpoint-wise paired $t$-test ($p \leq 0.023$; planetary and synoptic $p \leq 0.001$), with 8-9 of the ten checkpoints favouring ZWD in every band. \textit{Total} is LSD integrated over the full wavenumber range, not the sum of the band values.}
\label{tab:app_precip_lsd}
\small
\setlength{\tabcolsep}{8pt}
\begin{tabular}{lccc}
\toprule
Band & Without ZWD & With ZWD & Rel.\ $\Delta$ \\
\midrule
Planetary        & $0.191 \pm 0.100$ & $\mathbf{0.115 \pm 0.078}$ & $-39.9\%$ \\
Synoptic         & $0.225 \pm 0.110$ & $\mathbf{0.150 \pm 0.080}$ & $-33.3\%$ \\
Upper mesoscale  & $0.607 \pm 0.120$ & $\mathbf{0.526 \pm 0.092}$ & $-13.4\%$ \\
Lower mesoscale  & $1.018 \pm 0.120$ & $\mathbf{0.961 \pm 0.100}$ & $-5.6\%$ \\
\midrule
\textit{Total}   & $0.934 \pm 0.120$ & $\mathbf{0.876 \pm 0.098}$ & $-6.2\%$ \\
\bottomrule
\end{tabular}
\end{table}

\subsection{Cost to the pretrained variables}

The accompanying cost of ZWD on the standard pretrained variables over the same ten checkpoints is reported in Supplementary Table~\ref{tab:app_precip_cost}. Only the two 10\,m wind components degrade significantly on MAE, and by only ${\sim}1\%$; 2\,m temperature, mean sea-level pressure and specific humidity all fall within checkpoint-to-checkpoint noise, while the more large-error-sensitive RMSE additionally flags 2\,m temperature as a small ($+3.9\%$) significant degradation. All five variables retain Pearson $R \geq 0.997$. Across the ten checkpoints the ZWD model therefore improves precipitation consistently (by up to $+8.8\%$ in ETS at the 99th percentile, significant under the paired test), at a largely noise-level cost to the standard variables.

\begin{table}[h!]
\centering
\caption{Cost of adding ZWD to the standard pretrained variables and specific humidity, as a mean $\pm$ s.d.\ over the same ten training checkpoints. Columns give the RMSE for each model, then the relative change with two-sided paired $t$-test $p$-value for both RMSE and MAE (the MAE relative change and $p$-value equal those of the normalised MAE, the two differing only by a fixed per-variable scale factor). $\%\Delta>0$ means With ZWD has the larger error; $^{*}\,p<0.05$.}
\label{tab:app_precip_cost}
\small
\resizebox{\textwidth}{!}{%
\begin{tabular}{lccccc}
\toprule
& \multicolumn{2}{c}{RMSE (mean $\pm$ s.d.)} & \multicolumn{2}{c}{RMSE} & \multicolumn{1}{c}{MAE} \\
\cmidrule(lr){2-3}\cmidrule(lr){4-5}\cmidrule(lr){6-6}
Variable & Without ZWD & With ZWD & $\%\Delta$ & paired $p$ & $\%\Delta$ ($p$) \\
\midrule
u10 [m\,s$^{-1}$]        & $0.366 \pm 0.001$ & $0.371 \pm 0.001$ & $+1.3\%$ & $0.000^{*}$ & $+1.5\%$ ($0.000^{*}$) \\
v10 [m\,s$^{-1}$]        & $0.377 \pm 0.003$ & $0.381 \pm 0.001$ & $+1.1\%$ & $0.001^{*}$ & $+1.2\%$ ($0.005^{*}$) \\
t2m [K]                  & $0.382 \pm 0.009$ & $0.397 \pm 0.011$ & $+3.9\%$ & $0.022^{*}$ & $+4.8\%$ ($0.158$) \\
MSLP [Pa]                & $21.49 \pm 0.43$  & $21.90 \pm 0.56$  & $+1.9\%$ & $0.118$     & $+1.8\%$ ($0.248$) \\
$q$ [$10^{-4}$\,kg\,kg$^{-1}$] & $2.444 \pm 0.022$ & $2.449 \pm 0.040$ & $+0.2\%$ & $0.726$ & $+0.2\%$ ($0.837$) \\
\bottomrule
\end{tabular}
}
\end{table}

\suppnote{Rollout skill as a function of lead time}
\label{app:rollout}

Headline metrics in the main text are quoted at a single 6-hour lead time, and the lead-time dependence of the threshold skill is given there as Fig.~3. This note documents the rollout in further detail. The two precipitation models are evaluated over autoregressive rollouts out to a lead time of 114\,h ($\approx$4.75\,days) at 6-hourly steps. Neither model received any rollout (multi-step) fine-tuning: the forecasts are produced purely autoregressively by feeding each 6-hour prediction back as the input for the next step, so the curves below reflect the intrinsic rollout stability of the 6-hour models. Scores are averaged over a set of 10 initialisation dates spanning the test set.

\begin{figure}[htbp]
\centering
\begin{subfigure}{0.49\linewidth}
\centering
\includegraphics[width=\linewidth]{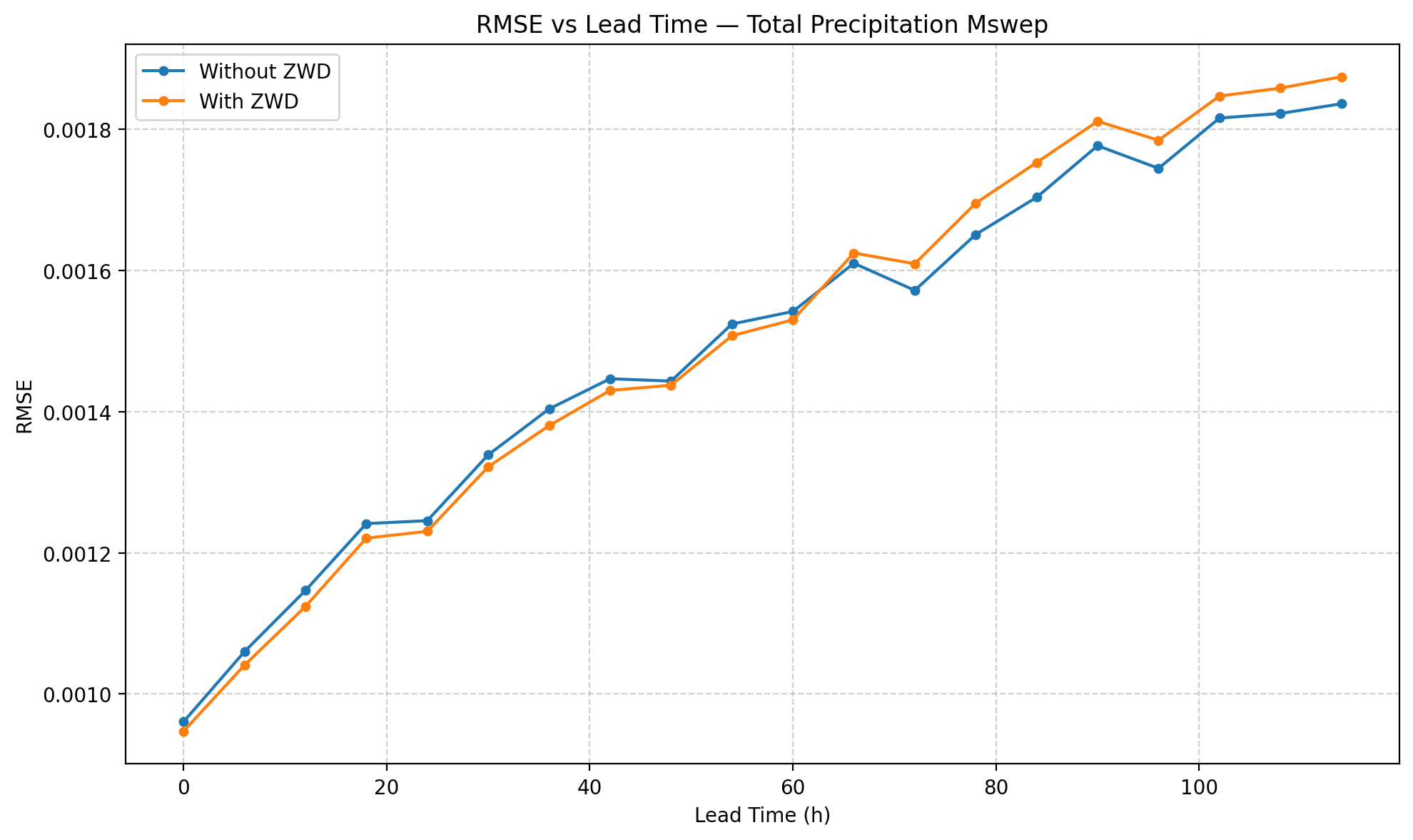}
\caption{Total precipitation.}
\label{fig:app_rollout_rmse_precip}
\end{subfigure}
\hfill
\begin{subfigure}{0.49\linewidth}
\centering
\includegraphics[width=\linewidth]{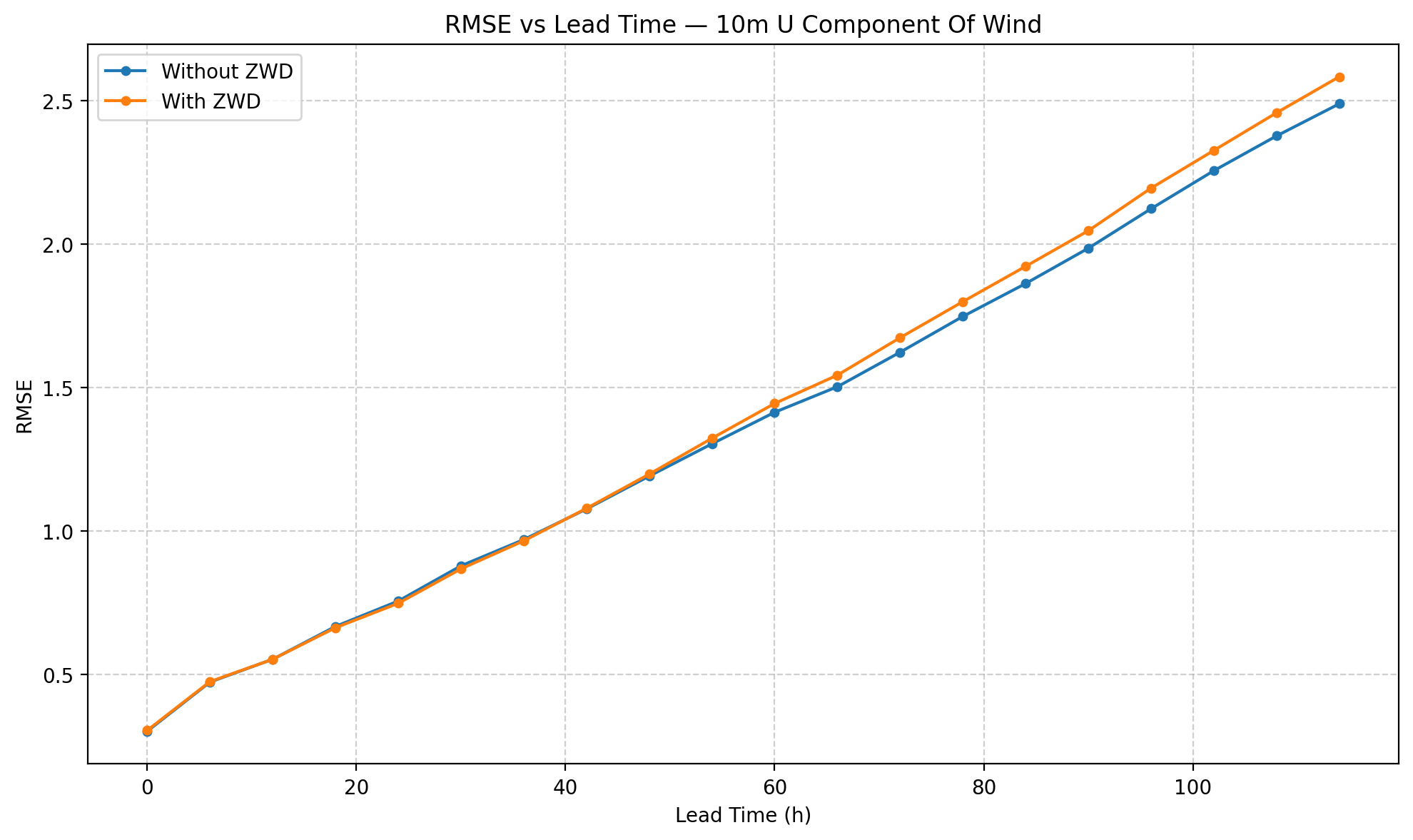}
\caption{10\,m zonal wind.}
\label{fig:app_rollout_rmse_u10}
\end{subfigure}
\caption{\textbf{RMSE as a function of lead time}, baseline (without ZWD) versus With ZWD, for \textbf{a} total precipitation and \textbf{b} the 10\,m zonal wind. Error grows smoothly over the rollout for both models, and the bulk-RMSE difference between them stays negligible throughout, confirming that adding ZWD does not degrade deterministic skill.}
\label{fig:app_rollout_rmse}
\end{figure}

We first verify that adding the ZWD objective does not compromise rollout stability, for precipitation or for the pretrained variables. Supplementary Fig.~\ref{fig:app_rollout_rmse} shows deterministic RMSE over the rollout for precipitation and for the 10\,m zonal wind. For both variables RMSE grows steadily and smoothly with lead time for both models, with no divergence. Consistent with the 6-hour analysis, the two models differ only negligibly in bulk RMSE: the With-ZWD model is marginally better than the baseline at the shortest precipitation leads and marginally worse at the longest, with the difference between the curves remaining negligible throughout.

This well-behaved decay extends to threshold-based skill, shown as ETS versus lead time at all four climatological percentiles in Fig.~3 of the main text. Two points there are worth stating explicitly. First, skill is lower in absolute terms at the more extreme thresholds: the 99th-percentile curve falls from $\approx$0.43 at the analysis time to $\approx$0.05 at 114\,h, against $\approx$0.70 to $\approx$0.26 at the 75th percentile. Second, beyond $\approx$70\,h the With- and Without-ZWD curves converge as both models lose skill, occasionally crossing within the noise, so the ZWD advantage is concentrated at early-to-intermediate lead times. Both models remain stable over the full 114\,h horizon at every threshold.

\suppnote{Temporal consistency and learning-quality examples}
\label{app:qualitative}

This note complements the aggregate metrics with temporally resolved skill and single-case spatial illustrations for the 1.3\,B model.

\begin{figure}[htbp]
\centering
\includegraphics[width=\linewidth]{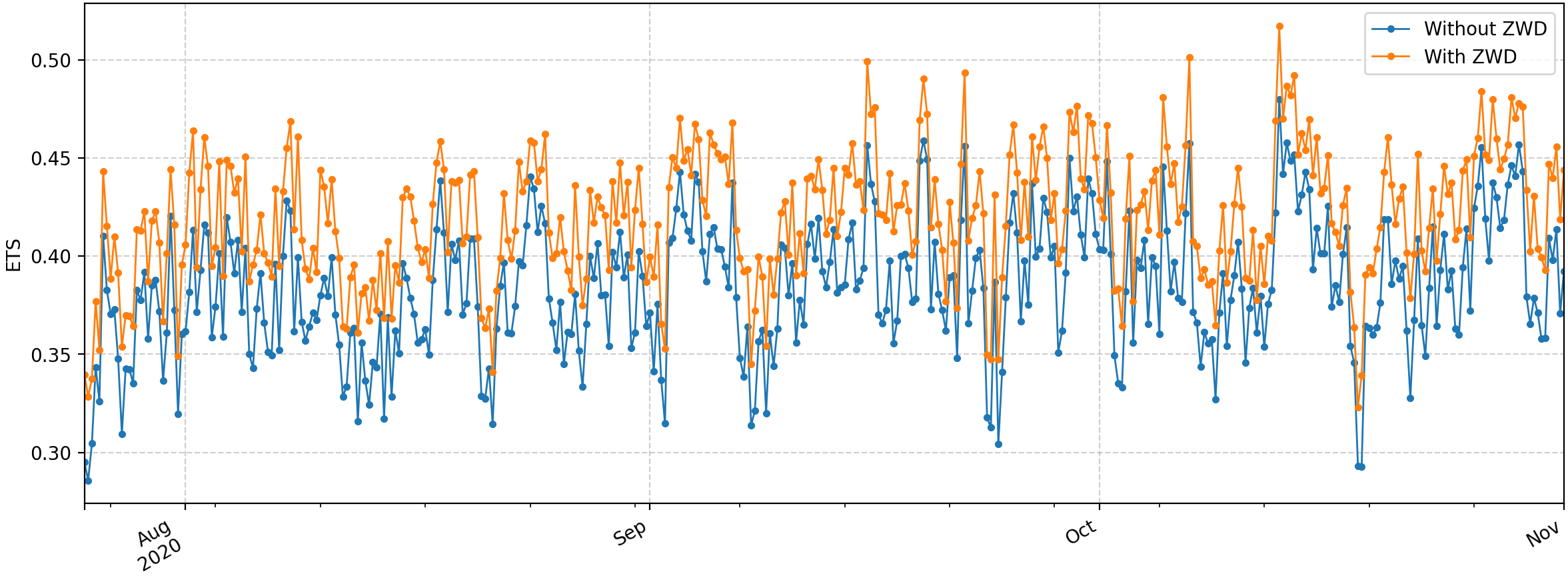}
\caption{\textbf{ETS at the 99th percentile over the test period for the 1.3\,B Aurora}, Model~A (``With ZWD'') versus Model~B (``Without ZWD''). The ZWD model leads at essentially every timestep.}
\label{fig:app_ets99_ts}
\end{figure}

Supplementary Fig.~\ref{fig:app_ets99_ts} shows the Equitable Threat Score at the 99th percentile over the test period for Models~A and~B. The ZWD-enriched model outperforms the baseline at essentially every timestep, establishing that the aggregate threshold gain reported in Table~3 of the main text is a robust property of the trained model rather than the product of a small number of outlier events.

\begin{figure}[htbp]
\centering
\includegraphics[width=\linewidth]{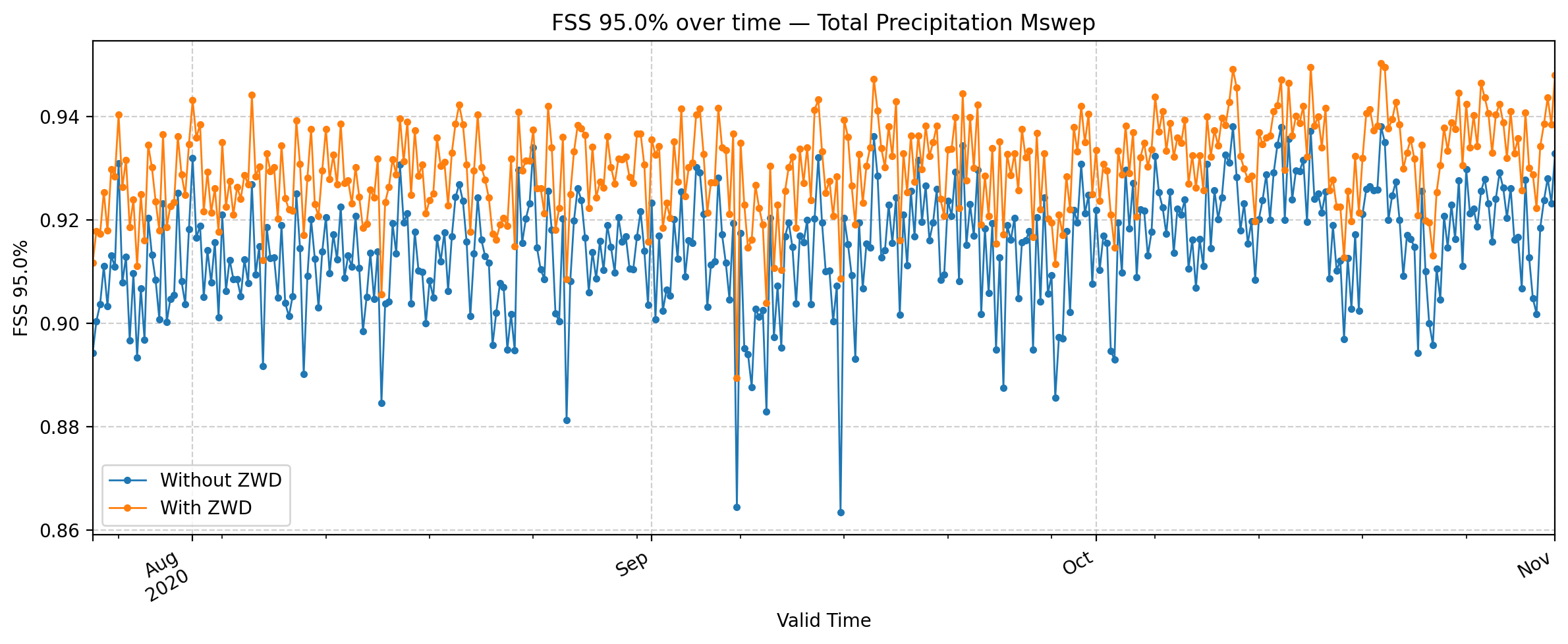}
\caption{\textbf{FSS at the 95th percentile over the test period for the 1.3\,B Aurora}, baseline (without ZWD) versus With ZWD.}
\label{fig:app_fss_ts}
\end{figure}

Supplementary Fig.~\ref{fig:app_fss_ts} shows FSS at the 95th percentile over the full test period for Models~A and B. The ZWD model lies above the baseline at nearly every verification time, confirming that the aggregate FSS improvement reported in the main text reflects a temporally consistent gain rather than a few favourable dates.

\begin{figure}[htbp]
\centering
\includegraphics[width=\linewidth]{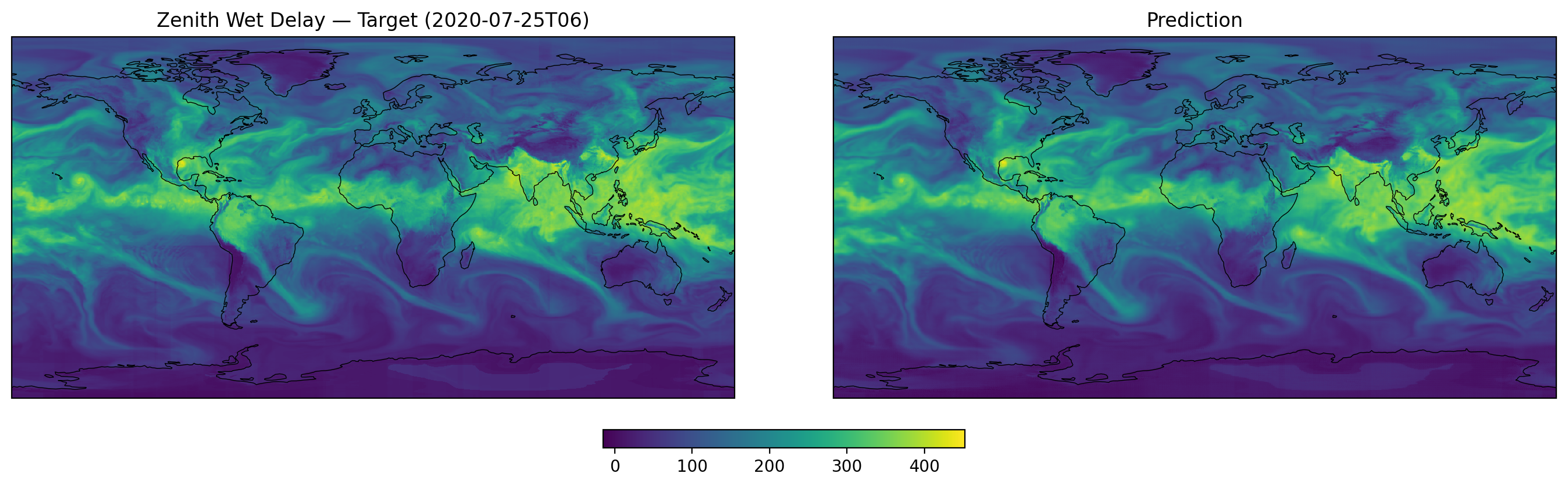}
\caption{\textbf{Step~1 ZWD target (left) and Aurora prediction (right)} when initialised on 2020-07-25T06.}
\label{fig:qual_zwd}
\end{figure}

\begin{figure}[p]
\centering
\includegraphics[width=0.62\linewidth]{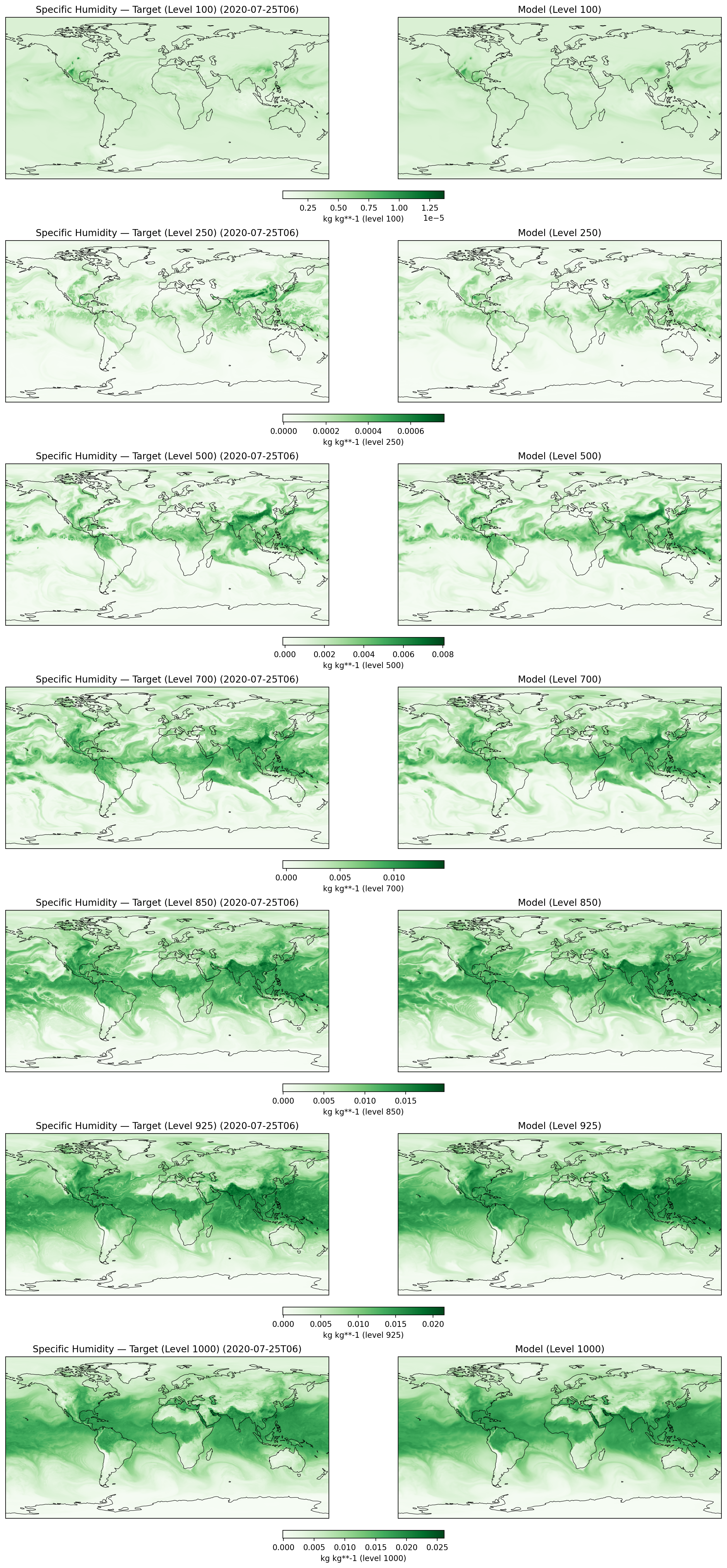}
\caption{\textbf{Specific humidity target (left) and Aurora prediction (right)} across pressure levels when initialised on 2020-07-25T06.}
\label{fig:qual_q}
\end{figure}

Supplementary Figs.~\ref{fig:qual_zwd} and~\ref{fig:qual_q} illustrate the spatial quality of the Step~1 prediction for a representative initialisation. ZWD (Supplementary Fig.~\ref{fig:qual_zwd}) is a variable outside Aurora's original vocabulary, yet the prediction reproduces its observed large-scale spatial structure and regional moisture gradients, rather than merely matching it in the bulk metrics of Table~1 of the main text. Specific humidity (Supplementary Fig.~\ref{fig:qual_q}), an already-pretrained moisture field, is shown alongside as a reference: the newly learned ZWD attains a visual fidelity comparable to this established variable, the spatial-domain counterpart to the near-identical convergence of their training curves (Supplementary Note~1). Qualitative precipitation forecasts comparing target, baseline and With-ZWD models are shown for three individual high-impact events as Fig.~4 of the main text, with further per-event diagnostics in Supplementary Note~8.

The spectral skill of the Step~1 ZWD prediction is reported in Supplementary Table~\ref{tab:zwd_skill_lsd}. The integrated LSD across all wavenumber bands is well within the envelope of the pretrained variables and better than that of u10, with the closest spectral match at planetary and synoptic scales. This complements the deterministic and threshold-based skill given in Table~1 of the main text.

\begin{table}[htbp]
\centering
\caption{Step~1 Log Spectral Distance (LSD) of ZWD across four wavenumber bands, compared to the range spanned by Aurora's standard pretrained variables; lower values indicate a more faithful power spectrum. The \textit{Total} row is the LSD integrated over the full wavenumber range, not the sum of the band-wise values.}
\label{tab:zwd_skill_lsd}
\small
\setlength{\tabcolsep}{6pt}
\begin{tabular}{lccc}
\toprule
Band & Scale range (km) & LSD (ZWD) & LSD range across pretrained variables \\
\midrule
Planetary        & $5{,}000$--$20{,}000$ & 0.006 & 0.005 (MSLP)--0.008 (t2m) \\
Synoptic         & $1{,}000$--$5{,}000$ & 0.022 & 0.015 (v10)--0.021 (u10) \\
Upper mesoscale  & $250$--$1{,}000$ & 0.125 & 0.045 (MSLP)--0.130 (u10) \\
Lower mesoscale  & $55$--$250$ & 0.279 & 0.126 (MSLP)--0.379 (u10) \\
\midrule
Total   & $55$--$20{,}000$ & 0.252 & 0.113 (MSLP)--0.339 (u10) \\
\bottomrule
\end{tabular}
\end{table}

\suppnote{Regional extreme-event case studies}
\label{app:regional}

The aggregate metrics in the main text are computed globally over the held-out test period. To address whether the ZWD-induced improvement also manifests during individual high-impact precipitation events (and to provide the region-aggregated time series around heavy-rain episodes that complement the global diagnostics) we examine three extreme events drawn from distinct basins and meteorological regimes, all falling within the test set:

\begin{itemize}
\item \textbf{South Asia monsoon} (initialised 2020-08-04 12Z): an active-monsoon episode producing extreme accumulations over the Indian subcontinent and the Bay of Bengal.
\item \textbf{East Asia Typhoon Bavi} (initialised 2020-08-24 06Z): a mature typhoon tracking towards the Korean peninsula.
\item \textbf{Central America Hurricane Delta} (initialised 2020-10-04 06Z): a rapidly intensifying hurricane over the north-western Caribbean.
\end{itemize}

For each event we evaluate the 1.3\,B Models~A (``With ZWD'') and~B (``Without ZWD'') over a regional domain enclosing the system. Two diagnostics are given in the main text: the per-cell \textit{skill difference} over extreme cells (the absolute ``Without ZWD'' error minus the absolute ``With ZWD'' error, so that green indicates the ``With ZWD'' model is closer to the observation) as Fig.~4, and the \textit{skill gain by intensity bin} - which stratifies the mean error reduction by the observed-precipitation percentile of each grid cell - as Fig.~5; across all three events the latter shows the same signature, ZWD being neutral or marginally negative at low-to-moderate intensities and strongly positive in the extreme upper tail. This note adds, for each event: the 6-hour accumulated precipitation forecasts against the MSWEP target; the regional spatial RMSE and percentile-threshold ETS as time series spanning ${\pm}90$\,h around the event; and the threshold-exceedance hit/miss/false-alarm classification.

\subsection{South Asia monsoon (2020-08-04)}

This is the clearest of the three cases. The regional RMSE time series (Supplementary Fig.~\ref{fig:reg_sa_rmse}) shows the ``With ZWD'' model below the baseline ``Without ZWD'' at nearly every 6-hourly step across the full ${\pm}90$\,h window, with the largest separation at the high-RMSE peaks; the threshold ETS series (Supplementary Fig.~\ref{fig:reg_sa_ets}) shows a consistent ``With ZWD'' advantage that grows with threshold, most pronounced at the $\geq$10 and $\geq$20\,mm levels. The skill-by-intensity decomposition (Fig.~5a of the main text) rises monotonically with intensity to $+1.52$\,mm of mean error reduction in the top percentile bin, while the spatial skill-difference map (Fig.~4a of the main text) is overwhelmingly green over the regions of heaviest rainfall, and the exceedance panels (Supplementary Fig.~\ref{fig:reg_sa_exceed}) show fewer missed events for the With-ZWD model at the higher thresholds.

\begin{figure}[htbp]
\centering
\includegraphics[width=0.95\linewidth]{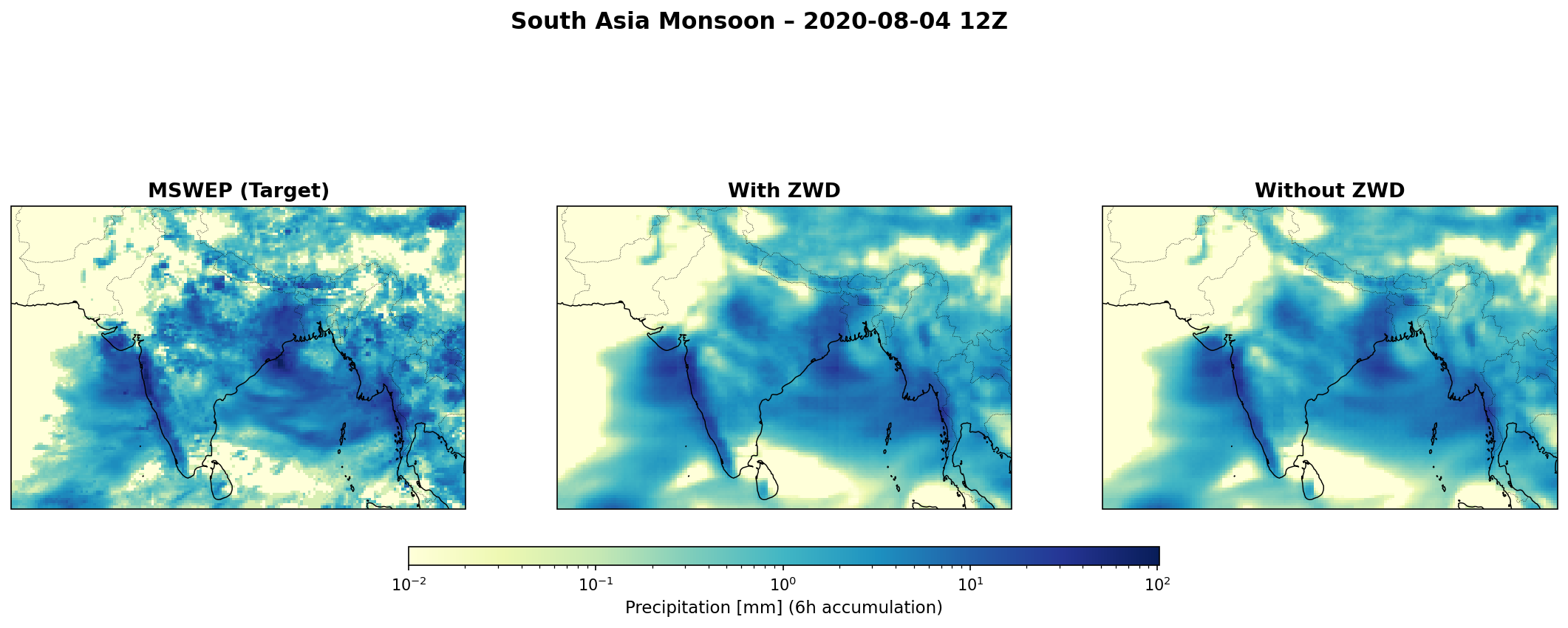}
\caption{\textbf{South Asia monsoon (2020-08-04 12Z):} 6-hour accumulated precipitation for the MSWEP target (left), model ``With ZWD'' (centre) and model ``Without ZWD'' (right).}
\label{fig:reg_sa_maps}
\end{figure}

\begin{figure}[htbp]
\centering
\begin{subfigure}{0.92\linewidth}
\centering
\includegraphics[width=\linewidth]{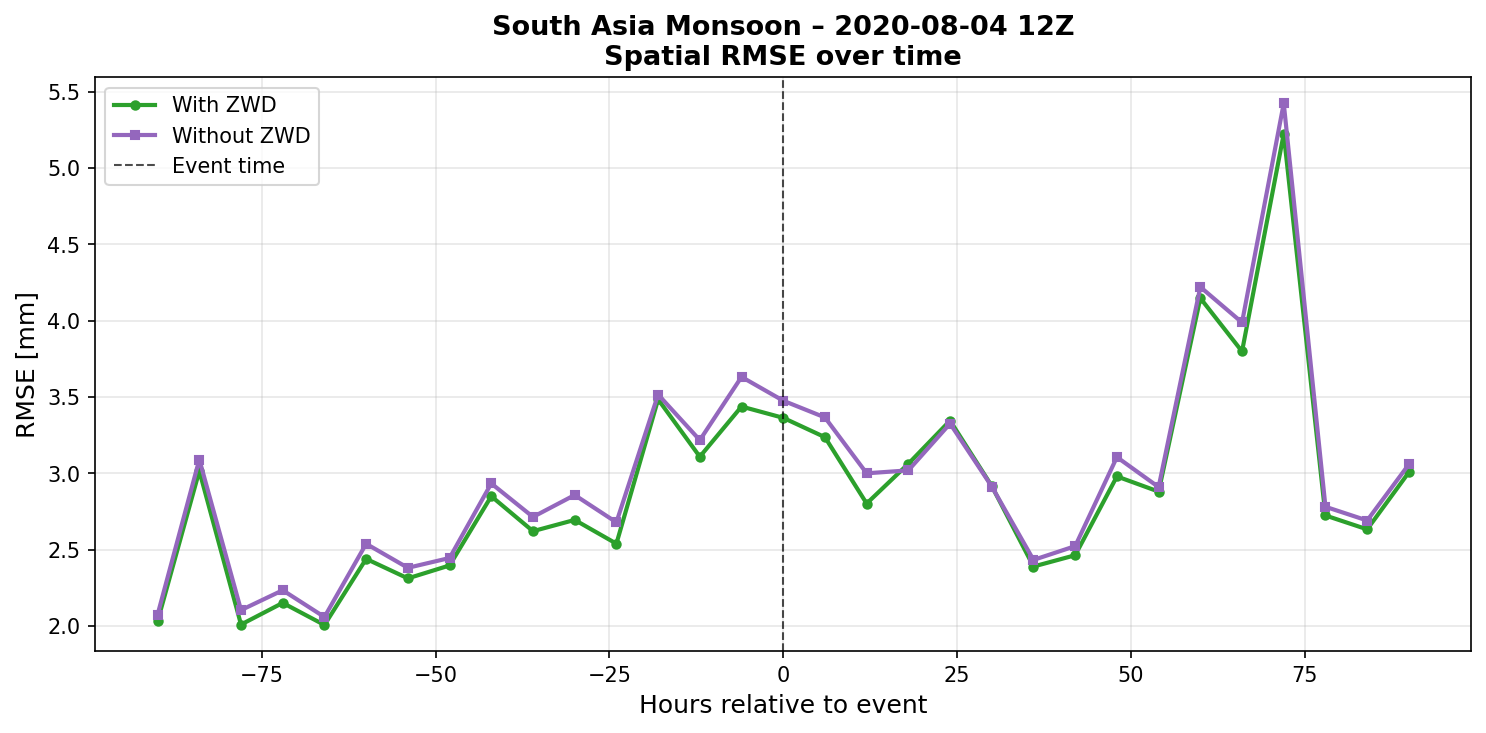}
\caption{Regional spatial RMSE.}
\label{fig:reg_sa_rmse}
\end{subfigure}

\vspace{0.4em}

\begin{subfigure}{0.92\linewidth}
\centering
\includegraphics[width=\linewidth]{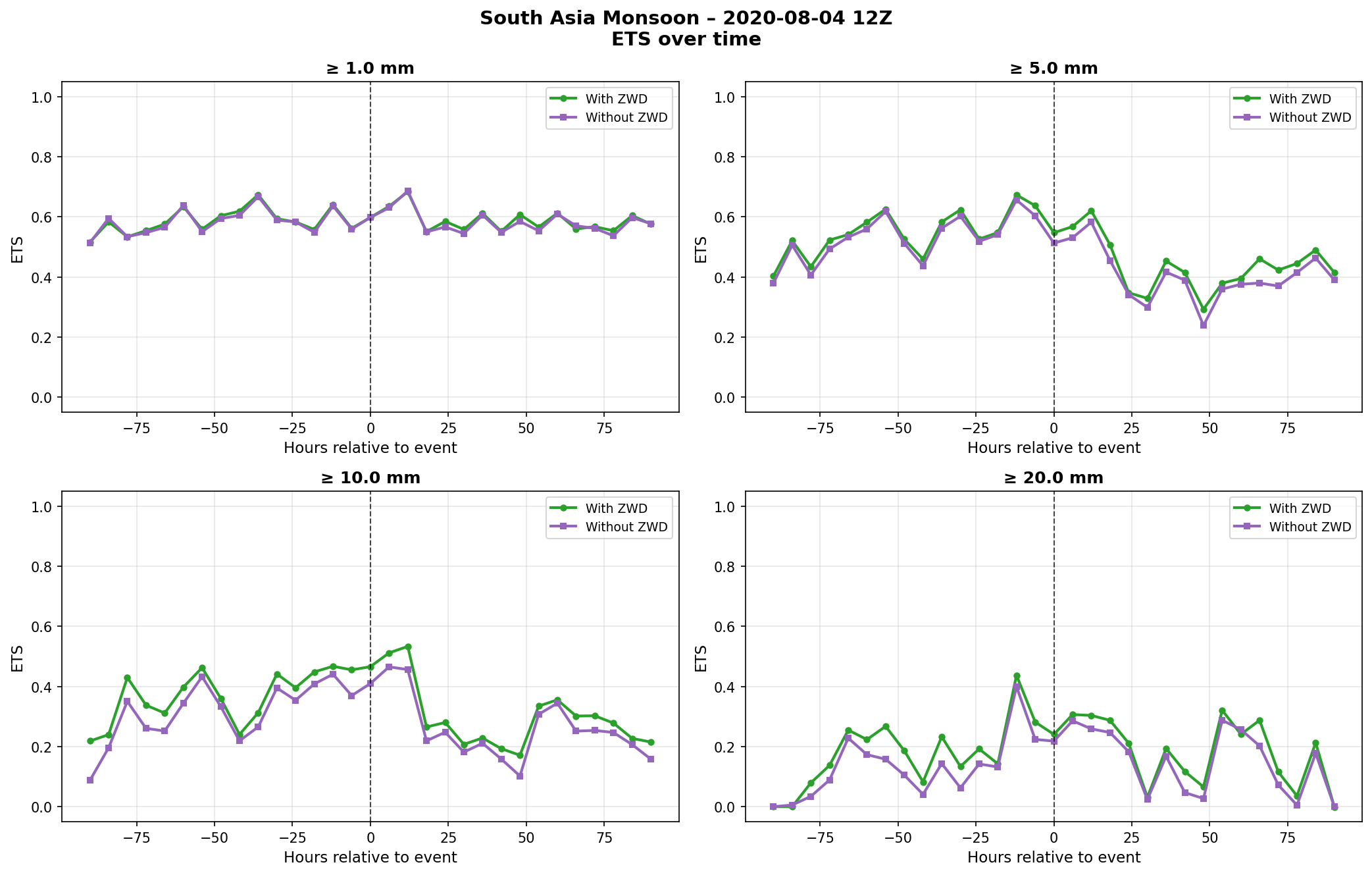}
\caption{Regional ETS at four climatological thresholds.}
\label{fig:reg_sa_ets}
\end{subfigure}
\caption{\textbf{South Asia monsoon: region-aggregated skill as a function of time} relative to the event ($t=0$, dashed line), ``With ZWD'' versus ``Without ZWD''. The ``With ZWD'' model is consistently more skillful, most clearly at the heavier thresholds.}
\label{fig:reg_sa_ts}
\end{figure}

\begin{figure}[htbp]
\centering
\includegraphics[width=0.95\linewidth]{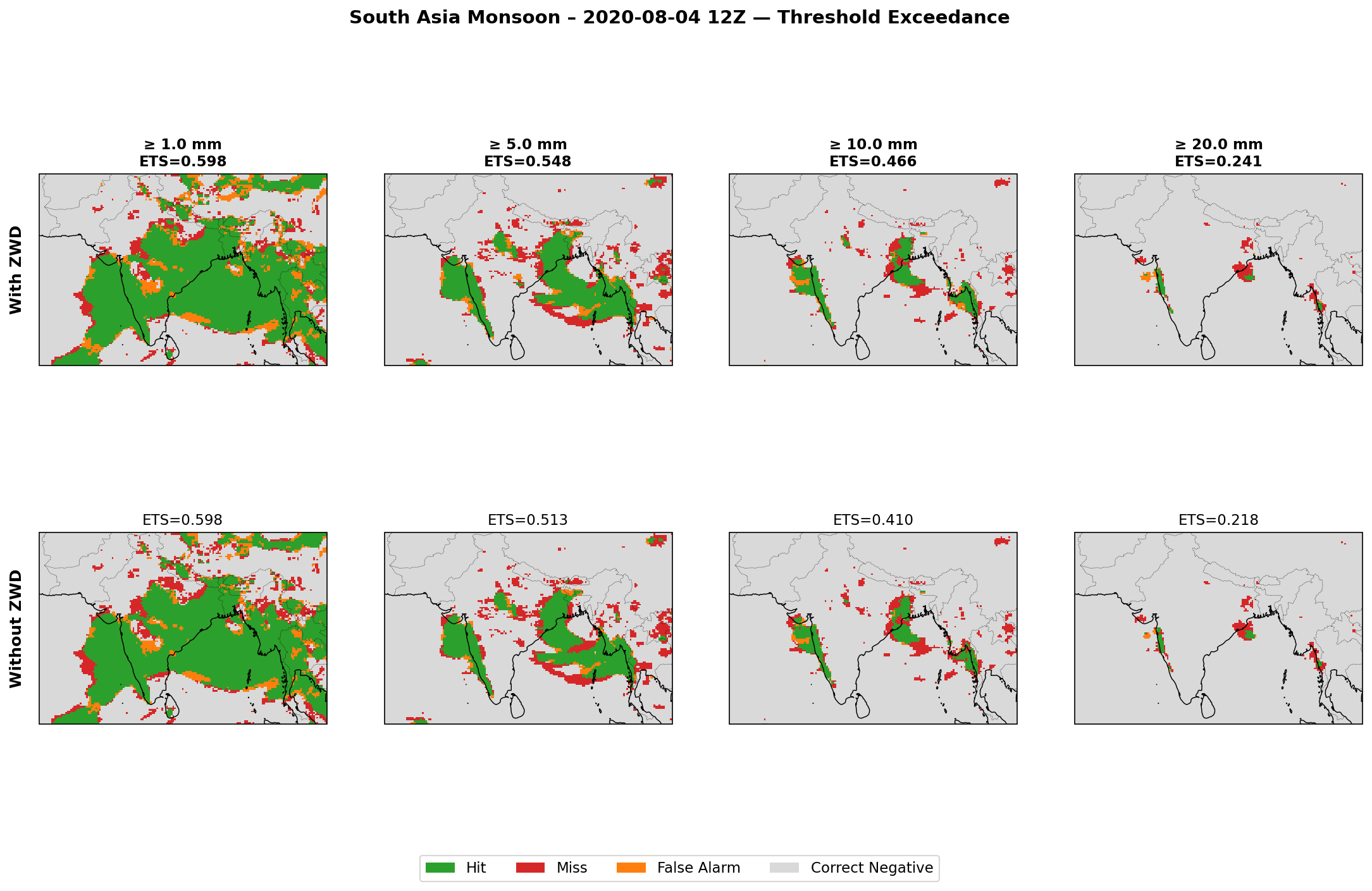}
\caption{\textbf{South Asia monsoon: threshold-exceedance classification} at four absolute thresholds (top: ``With ZWD''; bottom: ``Without ZWD''), giving the hit/miss/false-alarm/correct-negative breakdown behind the per-cell skill difference of Fig.~4a of the main text.}
\label{fig:reg_sa_exceed}
\end{figure}

\subsection{East Asia Typhoon Bavi (2020-08-24)}

The typhoon case confirms the same picture with a stronger extreme-tail signal. The skill-by-intensity decomposition (Fig.~5b of the main text) reaches $+2.17$\,mm of mean error reduction in the top percentile bin (the largest of the three events) and the spatial skill-difference map (Fig.~4b of the main text) shows a coherent green band of improvement along the typhoon's heavy-rain swath. The regional RMSE series (Supplementary Fig.~\ref{fig:reg_ea_rmse}) is noisier than the monsoon case, with the two models crossing at several steps, but the ``With ZWD'' model holds a clear advantage through the pre-event intensification phase and at the RMSE peak around the event time; the ETS series (Supplementary Fig.~\ref{fig:reg_ea_ets}) again favours the With-ZWD model most at the heavier thresholds.

\begin{figure}[htbp]
\centering
\includegraphics[width=0.95\linewidth]{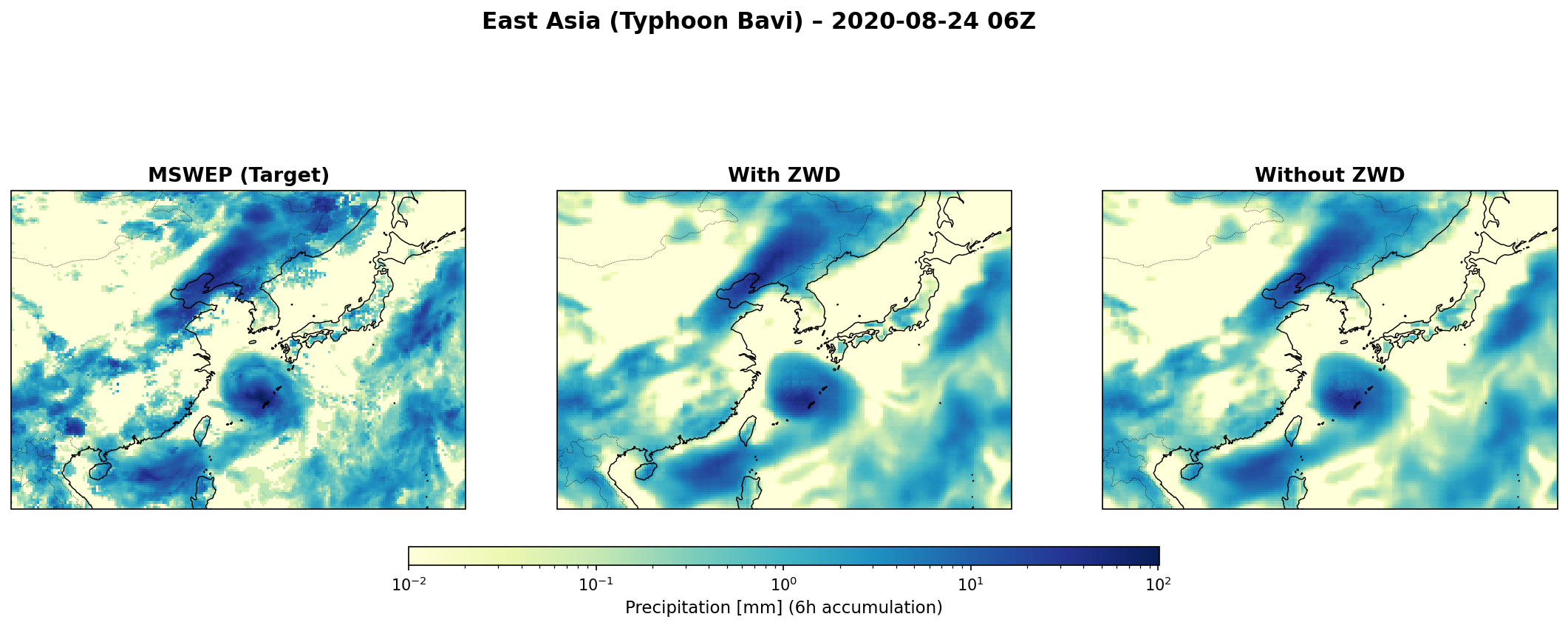}
\caption{\textbf{East Asia Typhoon Bavi (2020-08-24 06Z):} 6-hour accumulated precipitation for the MSWEP target (left), model ``With ZWD'' (centre) and model ``Without ZWD'' (right).}
\label{fig:reg_ea_maps}
\end{figure}

\begin{figure}[htbp]
\centering
\begin{subfigure}{0.92\linewidth}
\centering
\includegraphics[width=\linewidth]{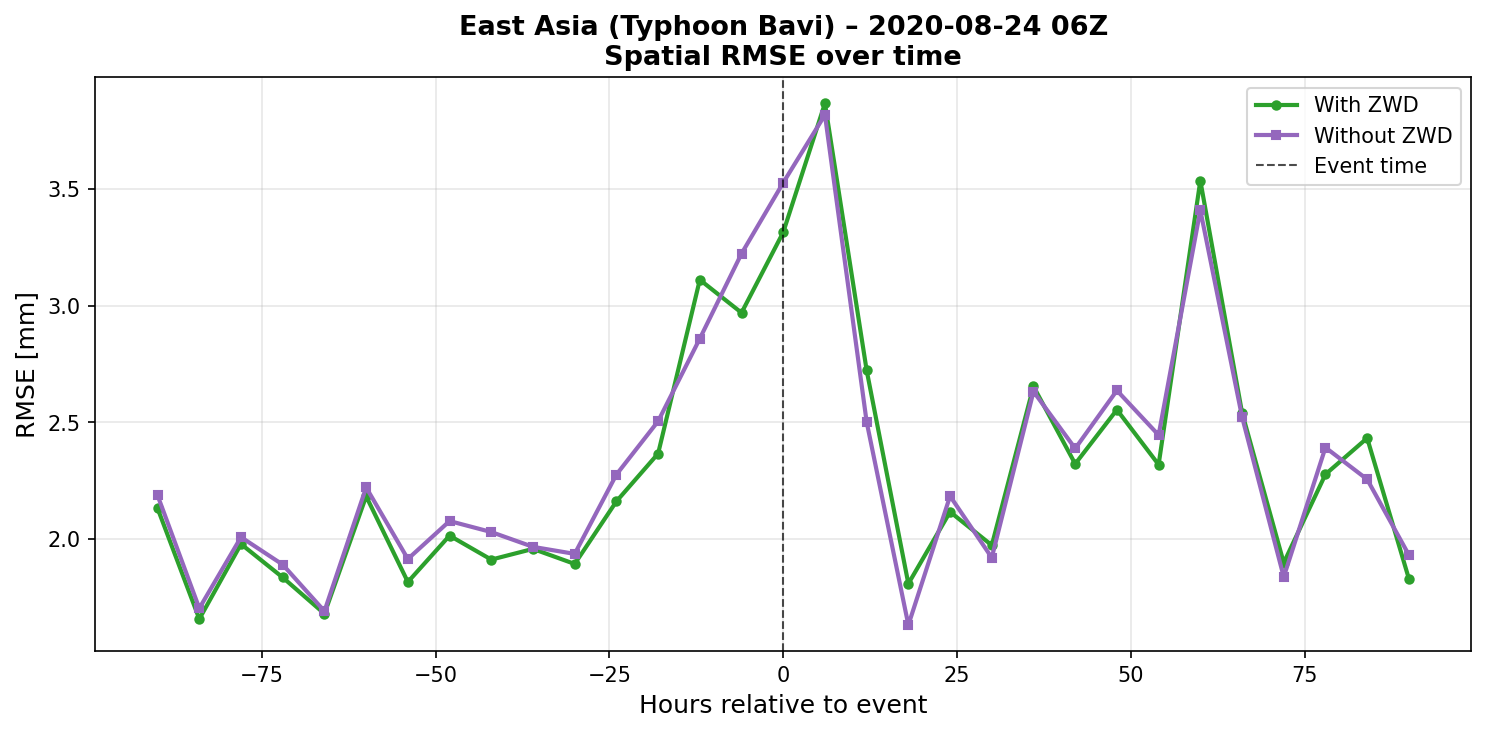}
\caption{Regional spatial RMSE.}
\label{fig:reg_ea_rmse}
\end{subfigure}

\vspace{0.4em}

\begin{subfigure}{0.92\linewidth}
\centering
\includegraphics[width=\linewidth]{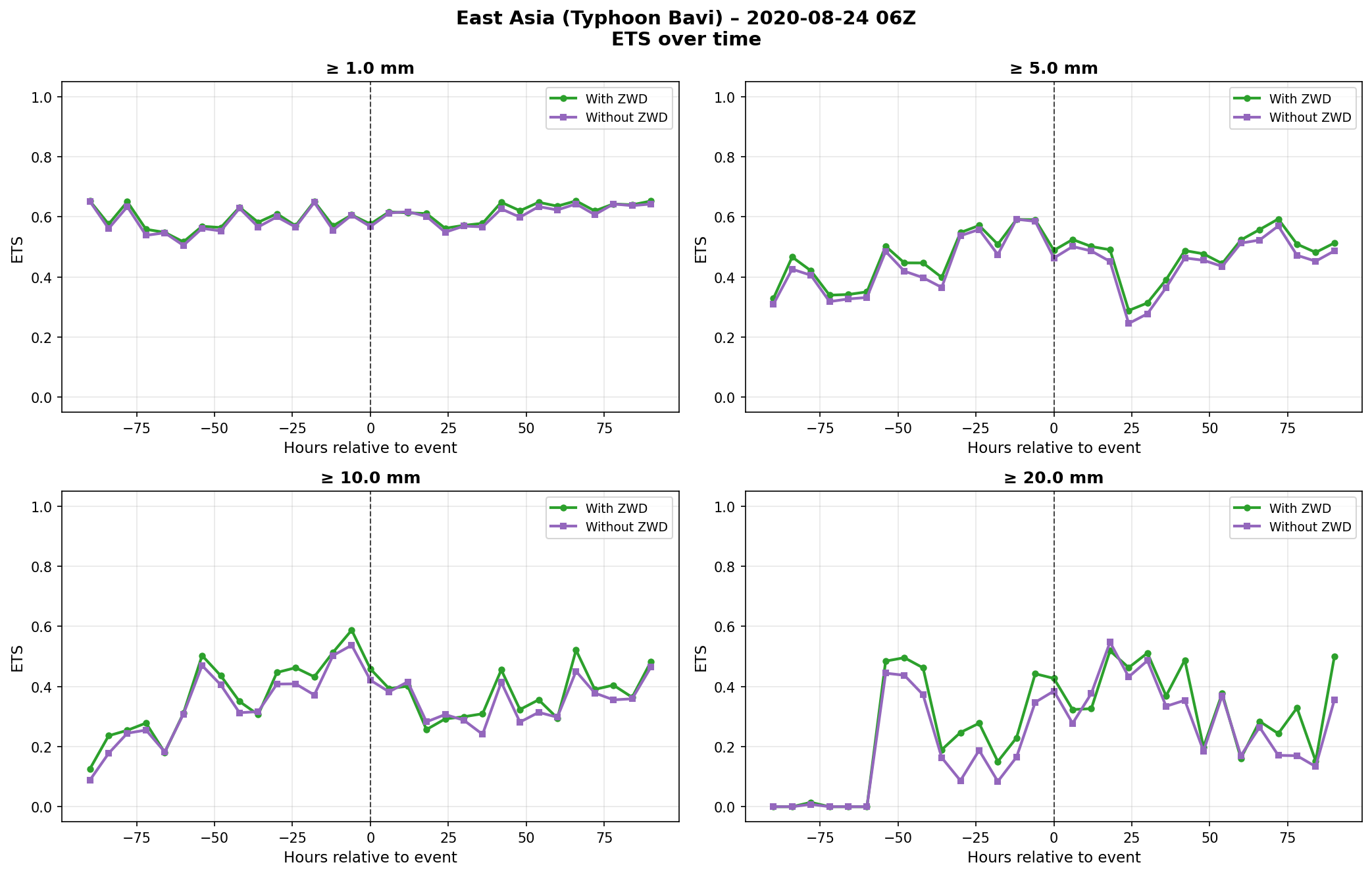}
\caption{Regional ETS at four climatological thresholds.}
\label{fig:reg_ea_ets}
\end{subfigure}
\caption{\textbf{East Asia Typhoon Bavi: region-aggregated skill as a function of time} relative to the event ($t=0$, dashed line), ``With ZWD'' versus ``Without ZWD''.}
\label{fig:reg_ea_ts}
\end{figure}

\begin{figure}[htbp]
\centering
\includegraphics[width=0.95\linewidth]{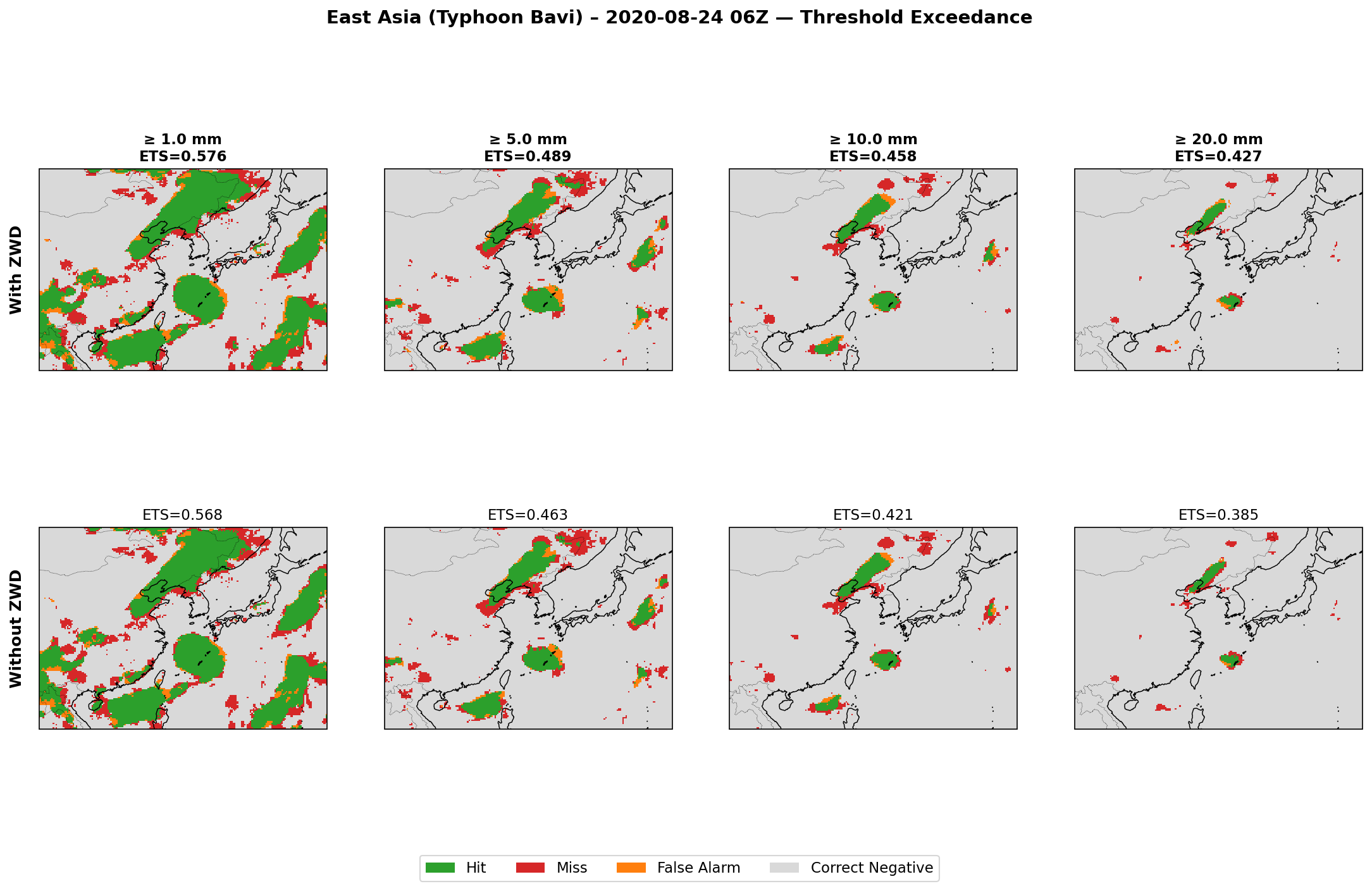}
\caption{\textbf{East Asia Typhoon Bavi: threshold-exceedance classification} at four absolute thresholds (top: ``With ZWD''; bottom: ``Without ZWD''), giving the hit/miss/false-alarm/correct-negative breakdown behind the per-cell skill difference of Fig.~4b of the main text.}
\label{fig:reg_ea_exceed}
\end{figure}

\subsection{Central America Hurricane Delta (2020-10-04)}

\begin{figure}[htbp]
\centering
\begin{subfigure}{0.95\linewidth}
\centering
\includegraphics[width=\linewidth]{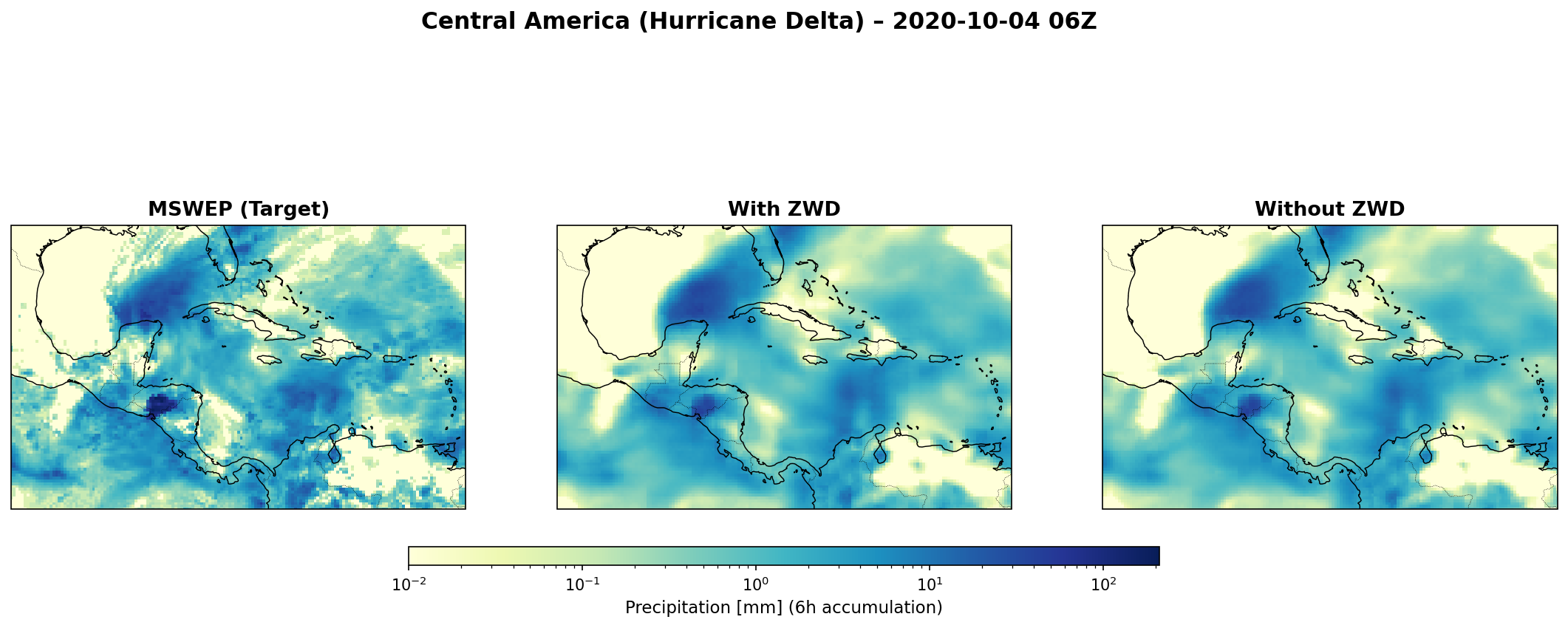}
\caption{6-hour accumulated precipitation: MSWEP target, ``With ZWD'', ``Without ZWD''.}
\label{fig:reg_ca_maps}
\end{subfigure}

\vspace{0.4em}

\begin{subfigure}{0.92\linewidth}
\centering
\includegraphics[width=\linewidth]{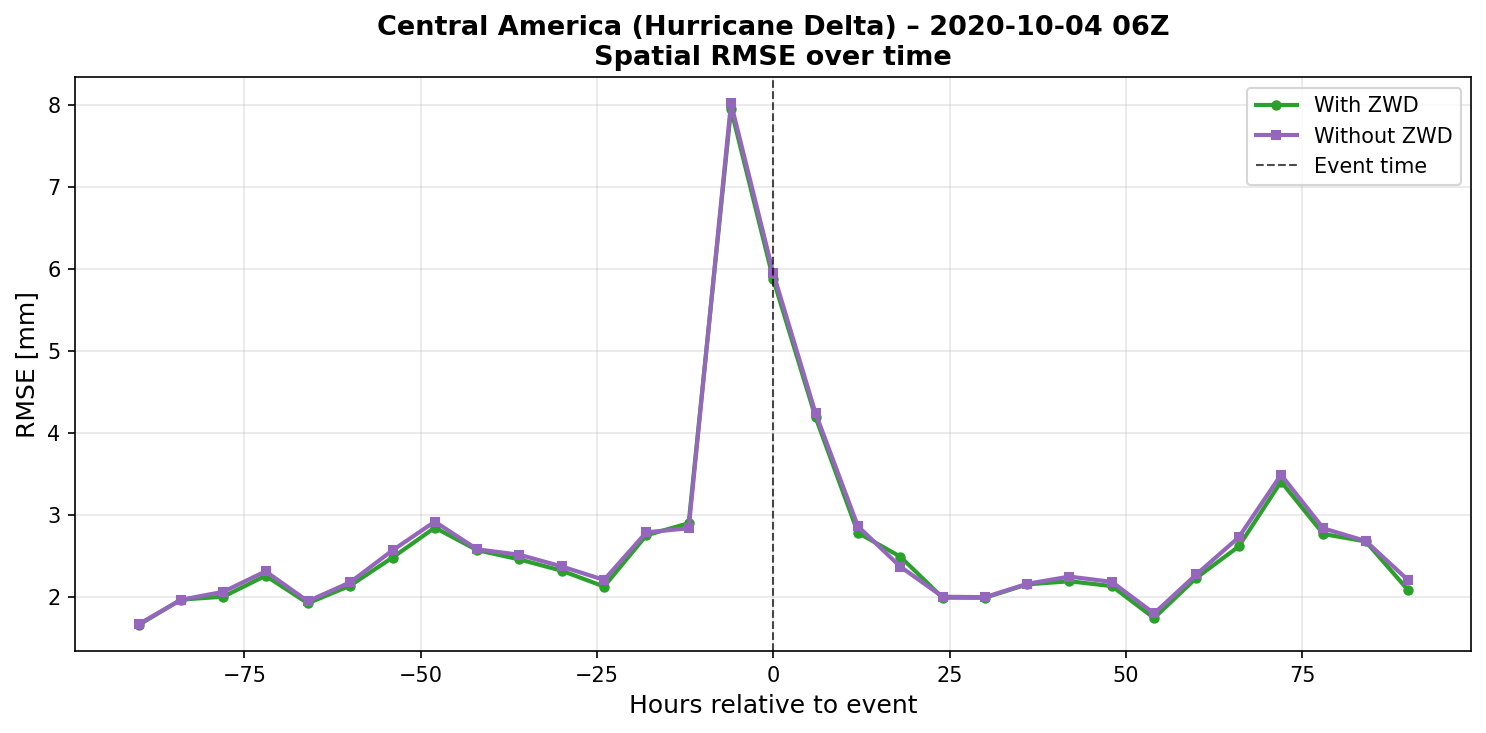}
\caption{Regional spatial RMSE over time.}
\label{fig:reg_ca_rmse}
\end{subfigure}

\caption{\textbf{Central America Hurricane Delta (2020-10-04 06Z): a more neutral case.} The bulk regional RMSE (\textbf{b}) is essentially unchanged between the two models, yet the extreme-tail skill gain persists (Fig.~5c of the main text), mirroring the other two events.}
\label{fig:reg_ca}
\end{figure}

The hurricane case is included as a more neutral counterexample. Here the region-aggregated RMSE series (Supplementary Fig.~\ref{fig:reg_ca_rmse}) shows the two models essentially overlapping, and the spatial skill-difference map (not shown) is mixed rather than predominantly green: adding ZWD neither helps nor harms the bulk forecast. Even so, the skill-by-intensity decomposition (Fig.~5c of the main text) preserves the same qualitative signature as the other two events, with a clear positive gain ($+0.71$\,mm) confined to the most extreme percentile bin. In all three cases the skill-by-intensity gain is thus confined to the most extreme bin, and the hurricane shows this tail gain with essentially unchanged bulk RMSE.

\end{document}